\documentclass[a4paper]{article}
\pdfoutput=1
\usepackage{pdfpages}

\usepackage{icrc2013}
\usepackage{amsmath}
\usepackage[english]{babel}
\usepackage{url}
\usepackage{hyperref}
\hypersetup{
  colorlinks=true,
  linkcolor=blue,
  citecolor=blue,
  urlcolor=blue,
}

\newcommand{\nosection}[1]{%
  \refstepcounter{section}%
  \addcontentsline{toc}{section}{\protect\numberline{\thesection}#1}%
  \markright{#1}}

\title{Observations of Cosmic Rays with HAWC: Contributions to ICRC 2013}

\authors{
{\bf The HAWC Collaboration:}\\
A.~U.~Abeysekara$^{a}$,
R.~Alfaro$^{b}$,
C.~Alvarez$^{c}$,
J.~D.~{\'A}lvarez$^{d}$,
R.~Arceo$^{c}$,
J.~C.~Arteaga-Vel{\'a}zquez$^{d}$,
H.~A.~Ayala Solares$^{e}$,
A.~S.~Barber$^{f}$,
B.~M.~Baughman$^{g}$,
N.~Bautista-Elivar$^{h}$
% ...
}

\afiliations{
{\center
$^{a}$ Department of Physics and Astronomy, Michigan State University, East Lansing, MI, USA\\
$^{b}$ Instituto de F{\'\i}sica, Universidad Nacional Aut{\'o}noma de M{\'e}xico, Mexico D.F., Mexico\\
$^{c}$ CEFyMAP, Universidad Aut{\'o}noma de Chiapas, Tuxtla Guti{\'e}rrez, Chiapas, Mexico\\
$^{d}$ Universidad Michoacana de San Nicol{\'a}s de Hidalgo, Morelia, Mexico\\
$^{e}$ Department of Physics, Michigan Technological University, Houghton, MI, USA\\
$^{f}$ Department of Physics and Astronomy, University of Utah, Salt Lake City, UT, USA\\
$^{g}$ Department of Physics, University of Maryland, College Park, MD, USA\\
$^{h}$ Universidad Politecnica de Pachuca, Pachuca, Hgo, Mexico\\
% ...
}
}

\abstract{
  We describe measurements of GeV and TeV cosmic rays with the High-Altitude
  Water Cherenkov Gamma-Ray Observatory, or HAWC, that were presented at the
  33$^\text{rd}$ International Cosmic Ray Conference (ICRC) in Rio de Janeiro,
  Brazil in July 2013.  The measurements include the observation of the shadow
  of the moon; the observation of small-scale and large-scale angular
  clustering of the TeV cosmic rays; the prospects for measurement of transient
  solar events with HAWC; and the observation of Forbush decreases with the
  HAWC engineering array and HAWC-30.
}


\begin{document}
\maketitle

\pagestyle{plain}
\pagenumbering{arabic}

\clearpage

\newpage
\onecolumn{
  \tableofcontents
}

\newpage
\setcounter{section}{0}
\nosection{Observation of the Moon Shadow and Characterization of the Point
Response of HAWC-30\\
{\footnotesize\sc Daniel Fiorino, Segev BenZvi, James Braun}}
\setcounter{section}{0}
\setcounter{figure}{0}
\setcounter{table}{0}
\setcounter{equation}{0}
%
%
%
%

\title{Observation of the Moon Shadow and Characterization of the Point Response of HAWC-30}

\shorttitle{HAWC Moon Shadow}

\authors{
Daniel W. Fiorino$^{1}$,
Segev BenZvi$^{1}$,
James Braun$^{2}$,
for the HAWC Collaboration.
}

\afiliations{
$^1$ WIPAC and Department of Physics, University of Wisconsin-Madison, Madison, WI \\
$^2$ Department of Physics, University of Maryland, College Park, MD \\
}

\email{dan.fiorino@icecube.wisc.edu}

\abstract{The High-Altitude Water Cherenkov (HAWC) Observatory is a TeV gamma-ray and cosmic-ray detector currently under construction at an altitude of 4100 meters at volcano Sierra Negra in the state of Puebla, Mexico.  Data taking has started during construction, and after nine months of operation, the air shower statistics are already sufficient to perform detailed studies of cosmic rays observed at the site. We report on the detection and study of the moon shadow and its deflection due to the geomagnetic field. From the observation of the moon shadow and simulations, we infer the pointing accuracy and the angular resolution of HAWC for cosmic rays.}


\maketitle

\section*{Introduction}
The HAWC Observatory is a second generation water Cherenkov observatory that is sensitive to cosmic-ray and gamma-ray induced air-showers of primary energies between 50 GeV and 100 TeV \cite{bib:mostafa}.  Water Cherenkov air-shower observatories are complimentary to pointing detectors like ground-based air Cherenkov detectors and spaced-based direct detection instruments. The near 100\% duty cycle and 2 sr instantaneous field of view of the HAWC Observatory provides an unbiased survey of the high-energy sky that triggers followup campaigns in air Cherenkov detectors, which have superior pointing resolution but small fields of view. Furthermore, the HAWC Observatory is continuing spectral measurements made by spaced-based direct detection instruments - like Fermi-LAT - because of its larger effective area at higher energies. The HAWC detector comprises 300 optically isolated water Cherenkov detectors (WCD) containing about 200,000 liters of filtered water and four upward-facing Hamamatsu photomultiplier tubes (PMTs). The modular design of HAWC makes data-taking possible during construction. Since September 2012, 30 WCDs (HAWC-30) have been operating, allowing us to gather useful diagnostic information through cosmic-ray observations. 

HAWC detects air showers through their cascading superluminal particles which penetrate the WCDs and then emit Cherenkov photons that strike the cathodes of PMTs causing a pulsed signal. Each PMT pulse is digitized at a central electronics house as a time over threshold (ToT) using custom front-end board electronics and then timestamped using a CAEN time-to-digital converter with ~100ps resolution.  In software, an air shower reconstruction is triggered if a simple multiplicity trigger condition is passed, that is \textit{N} PMT pulses in a time window \textit{T}. The ToT of each pulse is calibrated to an amount of charge deposited in the PMT \cite{bib:lauer}\cite{bib:ayala}. The spatial distribution of charge deposited on all PMTs within a given trigger window is used to find the shower core location - the original ground strike location of the primary particle. The core location and the calibrated relative PMT timing information \cite{bib:lauer}\cite{bib:ayala} is used to reconstruct the angle of incidence of the air shower.

Nearby objects, such as the Moon, cause detectable deficits on top of a nearly isotropic flux of incident air showers at Earth. Before the detection of the first gamma-ray source with HAWC-30, the observation of the cosmic-ray Moon shadow's angular width and position allows us to infer the resolution and pointing of the detector's angular reconstruction. As the experiment grows, this becomes a daily monitoring tool.

To produce a sky map of the statistical significance of the deficit in the vicinity of the position of the Moon, we compare a map of the actual cosmic-ray arrival directions to a reference map that represents the expected cosmic-ray flux in the absence of a Moon shadow.  The reference map is produced from the data themselves from methods \cite{bib:atkins} also used in point source searches.  The method has been adapted to use a HEALpix \cite{bib:gorski} binning which conserves angular area throughout the map. Angular bins approximately 0.1$^\circ$ in resolution were used. After the reference map is made, it is compared to the data map and the significance is calculated using the method of Li \& Ma \cite{bib:lima}. As expected, this results in a Gaussian significance distribution of width one for a source free (or sink free) dataset. Map smoothing is applied to study the significance of the shadow as a function of bin size and thus estimate the angular resolution of the detector.

A Moon-centered equatorial coordinate system is used for this study. Each reconstructed cosmic-ray air shower direction is given an azimuthal coordinate of its incident right ascension minus the current Moon right ascension and a polar coordinate of its incident declination minus the current Moon declination. These are denoted as $\Delta$RA and $\Delta$Dec, respectively.

We expect the Moon shadow position and width as well as the angular resolution of the detector to be a function of energy. HAWC-30 does not have an accurate energy reconstruction but we can use the number PMT channels hit in the air shower (nCh) as an energy proxy. For this study, the data are split into five equal-statistics bins in nCh. The highest nCh bin proves to provide the best reconstructions and highest Moon signal; this bin corresponds to nCh $\geq$ 32. From simulation we determine the median energy of this subset to be 2.9 TeV. As only proton cosmic rays are considered in the simulation, this is a lower bound. The energy resolutions of the bins are quite wide, but this energy proxy will be replaced with a proper energy estimation when the detector becomes larger.

\section*{Simulation}

The Earth's magnetic field will widen and shift the observed shadow width. From simulation \cite{bib:benzvi} the cosmic-ray deflection ($\Delta$) from the Moon to Earth as observed at the HAWC site is approximately linear with primary energy $E$:

\begin{equation}
 \Delta = 1.58^\circ \cdot Z \left( \frac{\mathrm{TeV}}{E} \right) .
\end{equation}

Using a parametric simulation of the HAWC detector we simulate the expected Moon shadow observation together with the effects of the geomagnetic field. An eight-particle representative composition (H, $^{4}$He, $^{12}$C, $^{16}$O, $^{20}$Ne, $^{24}$Mg, $^{28}$Si, $^{56}$Fe) is used as input to the incoming particle flux in abundances that correspond to the CREAM-2\cite{bib:cream2} measured fluxes. Simulated cosmic-ray air showers triggered the HAWC detector in accordance with its detector response in energy, angle, and shower distribution. Shower directions were smeared in accordance with the simulated point-spread function of HAWC. Since the HAWC detector has undergone construction and expansion during the data run, a detector configuration of 86 PMTs was chosen. The background model is set to match the Milagro observed large-scale anisotropy \cite{bib:milagro}, but this 10$^{-3}$ effect was determined to have no bearing on the Moon shadow. For more information on the simulation technique see \cite{bib:benzvi}.

\begin{figure}
\centering
  \includegraphics[width=0.5\textwidth]{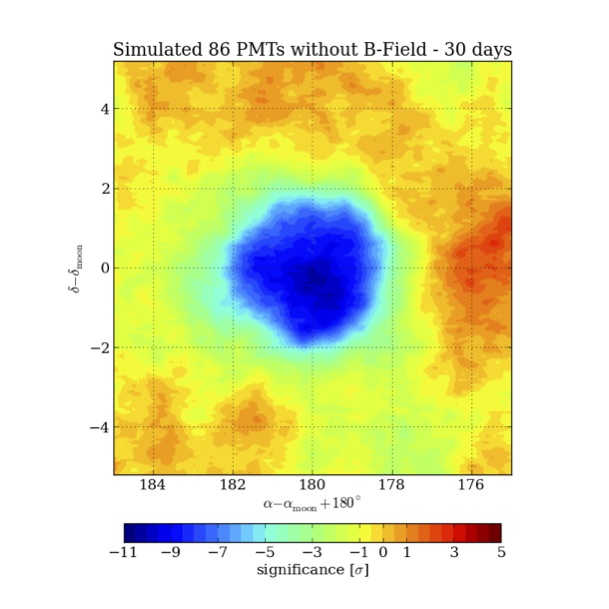}
  \includegraphics[width=0.5\textwidth]{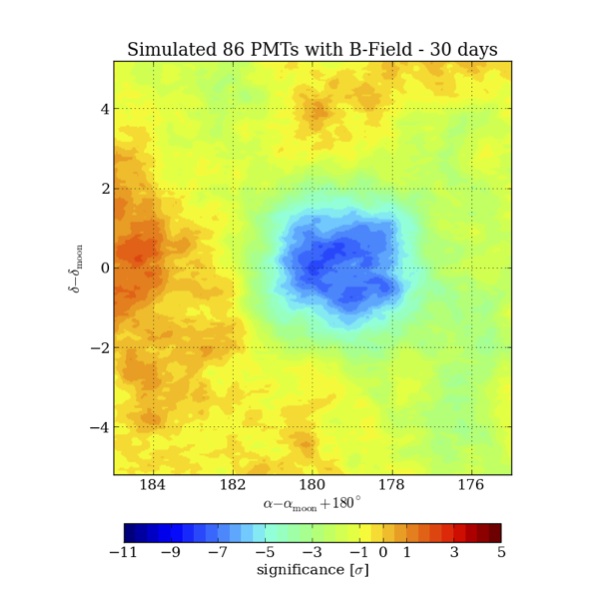}
  \caption{Skymap around the region of the Moon for 30 days of simulated HAWC-30 (86 PMTs)  data without (top) and with a geomagnetic field (bottom). Events were plotted in the bin of the difference of the air shower angle of incidence and the Moon position in equatorial coordinates. The plot shows the Li \& Ma significance. A cut of nCh $\geq$ 32  was applied.}
  \label{fig:geomag_sim_maps}
\end{figure}

\begin{figure}[t]
  \centering
  \includegraphics[width=0.5\textwidth]{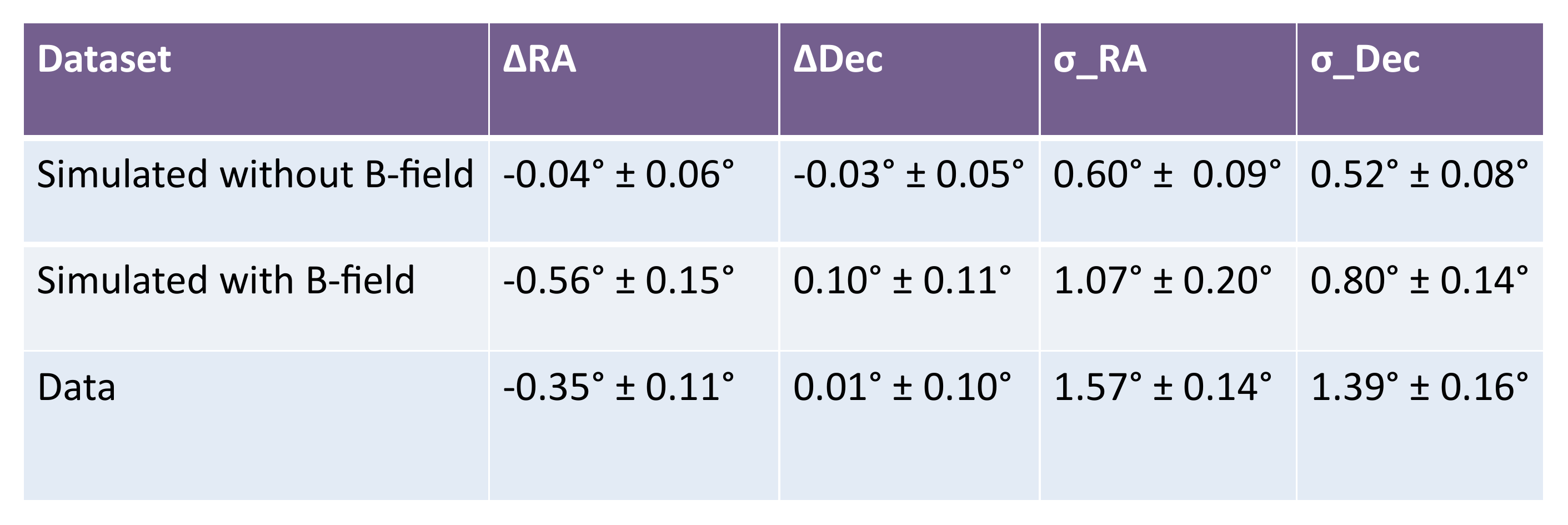}
  \caption{The results of an asymmetric two-dimensional Gaussian fit of the Moon shadow are shown for simulations with and without the EarthÕs geomagnetic field and HAWC-30 data. Notice that the width in the $\Delta$RA direction is larger than the width in $\Delta$Dec for data with a geomagnetic field. Also the simulation with the geomagnetic field  is consistent with data.}
  \label{fig:gaussfits_table}
\end{figure}

Using the IGRF 2011 \cite{bib:finlay}, we back-propagate simulated cosmic rays through the geomagnetic field to the radius of the Moon's orbit. If the cosmic ray collides with the disk of the Moon, we throw out the cosmic ray from the simulated dataset. This feature gives us the ability to study the widening and shift of the observed Moon shadow by comparing maps made with and without the geomagnetic field model. Figure~\ref{fig:geomag_sim_maps} shows the maps for 30 days of simulated HAWC-30 air showers around the Moon position with and without the geomagnetic field model. For nCh $\geq$ 32, a 37\% $\pm$ 16\% widening and offset in $\Delta$RA by 0.4$^\circ \pm$ 0.1$^\circ$ of the shadow is observed (Figure~\ref{fig:gaussfits_table}). The widening and shift diminishes with increasing median energy of the dataset. The center of the Moon shadow is determined from an asymmetric two-dimensional Gaussian fit to the relative intensity map. The center is shifted almost entirely in $\Delta$RA by an amount that depends upon the majority primary energy in the dataset.

 The $\Delta$RA shift in the Moon shadow center is energy-dependent and our dataset has a wide variance in energy, so we expect a shadow significance that drops off faster in $\Delta$RA than in $\Delta$Dec due to smearing and offsets in the shadow peaks as a function of energy. This is the reason. 

\section*{Angular Resolution of Detector Reconstruction}

It is helpful to use an independent estimate of the angular resolution of the detector reconstruction. One way to do this is by splitting the detector reconstruction into two sub-detectors so that each sub-detector covers the same physical area as the full detector but contains half as many PMTs. One way this is accomplished is by using every other PMT for one sub-detector and the remaining PMTs in the other. Even- and odd-numbered PMTs accomplish this end \cite{bib:alexandreas}. The difference in the angular reconstruction of the two sub-detectors ($\Delta_{EO}$) is plotted for all air showers. Assuming a Gaussian point-spread function of the detector, the median of this distribution ($M_{\Delta}$) is related to the angular resolution $\sigma$ of the full detector as follows:

\begin{equation}
 \sigma = \frac{M_{\Delta}}{2 \cdot 1.177}
\end{equation}

\begin{figure}[t]
  \centering
  \includegraphics[width=0.4\textwidth]{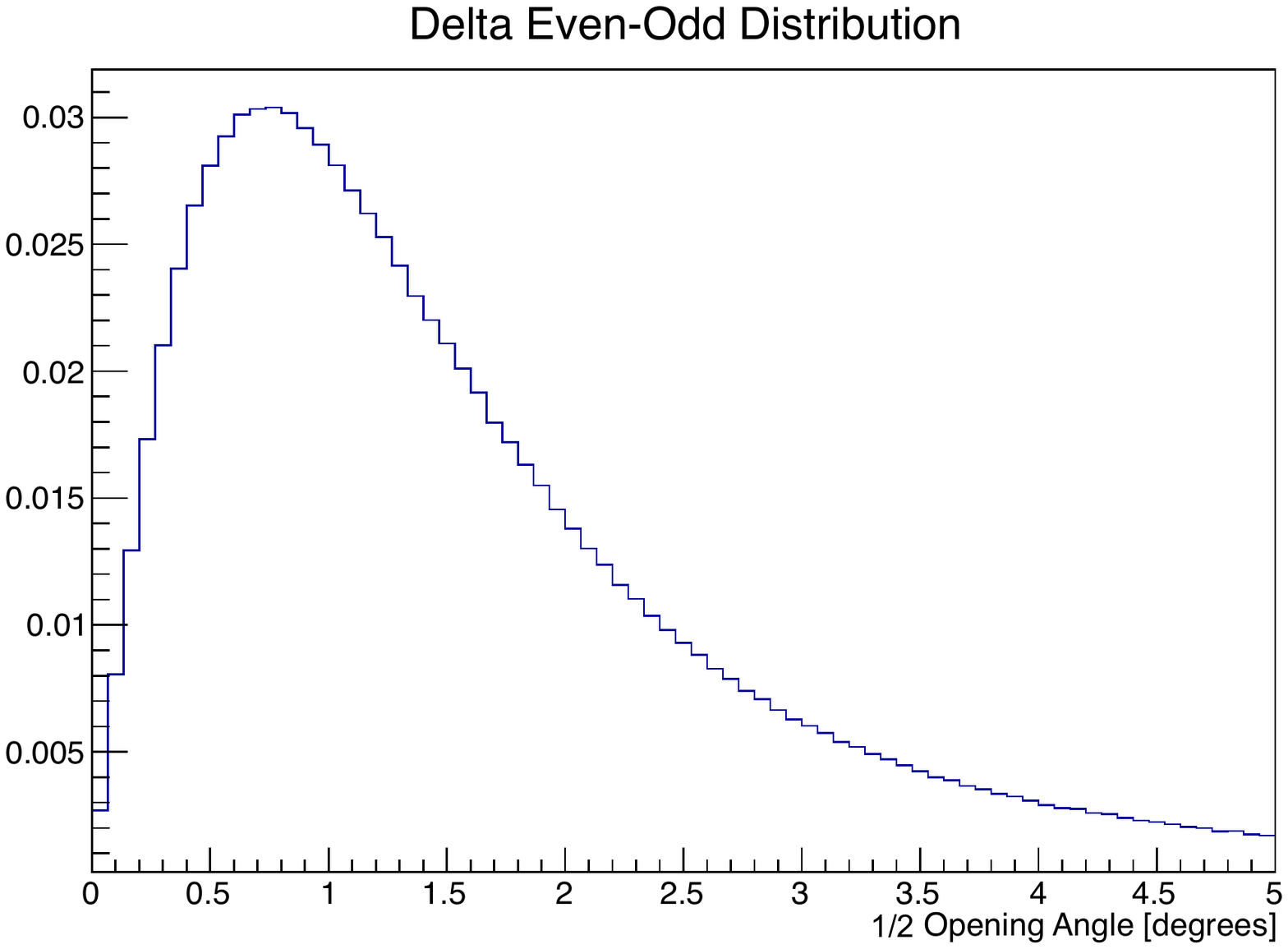}
  \caption{The normalized distribution of the half opening angle between the even-PMT and odd-PMT reconstructions for 100,000 air showers during a data run in March 2013 when HAWC had 102 PMTs. Data runs from earlier and later in the dataset were found to be consistent with this data run. The fitted width is 1.2$^\circ$, which is our estimated detector resolution.}
  \label{fig:deltaEO}
\end{figure}

The factor of 1.177 comes from the integration of the functional form of the distribution to the median. The factor of 2 is the result of two factors of $\sqrt2$, the first of which comes from adding the errors of each fit in quadrature. The second comes from the angular resolution scaling with $\sqrt{nPMTs}$ in the fit. For this reason, the distribution in Figure~\ref{fig:deltaEO} is the opening angle between the fits divided by 2. This method is not sensitive to systematic offsets in the detector pointing, which will be seen by both fits and cancel out in the opening angle distribution. 

The result for nCh $\geq$ 32 is a one-sigma width of 1.2$^\circ$ for the whole detector. Since the Moon's angular diameter is about 0.5$^\circ$, the Moon should appear as a Gaussian sink. Of course, the observed width of the moon is widened due to geomagnetic smearing. Again, from simulation we expect 37\% widening for the whole dataset.

\section*{Moon Shadow Observation}

We observed the Moon shadow from 2012 October 22 to 2013 April 11 and accumulated over 131 days of livetime. Using the data quality cut of nCh $\geq$ 32, seven billion air showers survive from this dataset. The peak significance is -15.6$\sigma$ and is centered at (179.6$^\circ \pm 0.1 ^\circ$, 0.0$^\circ \pm 0.1 ^\circ$) according to a two-dimensional Gaussian fit to the relative intensity map (Figure~\ref{fig:gaussfits_table}). Also according to the fit, the observed width in $\Delta$RA is 1.6$^\circ \pm$ 0.1 $^\circ$ which is consistent with our prediction of a 37\% $\pm$ 16\% widening of the point-spread function (1.2$^\circ$, see Figure~\ref{fig:deltaEO}). The shadow center is also consistent with the simulation results for the same energy bin. Furthermore, the blocked flux measurement in a circular region of radius 5$^\circ$ around the Moon shadow (0.255\%) matches the expected blocked flux from the Moon (0.250\%) from geometric considerations. 
\begin{figure}[t]
  \centering
  \includegraphics[width=0.5\textwidth]{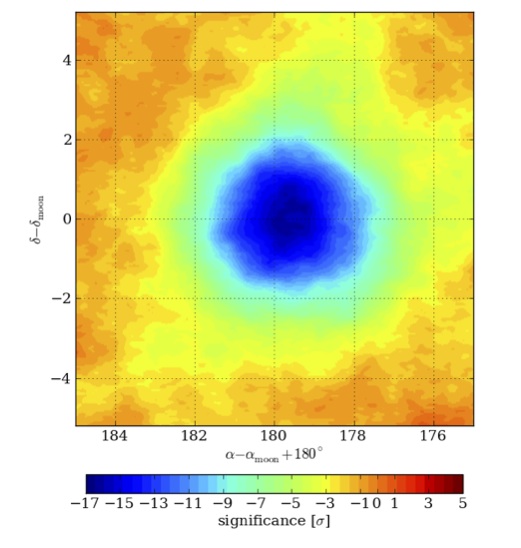}
  \caption{Skymap around the region of the Moon for 131 days of HAWC-30 data with detector configurations of 86 PMTs (50 days), 102 PMTs (53 days), and 114 PMTs (28 days). The same skymap technique is used in data and simulation. Background is estimated via Direct Integration\cite{bib:atkins}. A cut of nCh $\geq$ 32  was applied.}
  \label{data-moonmap}
\end{figure}

\begin{figure}[t]
  \centering
  \includegraphics[width=0.5\textwidth]{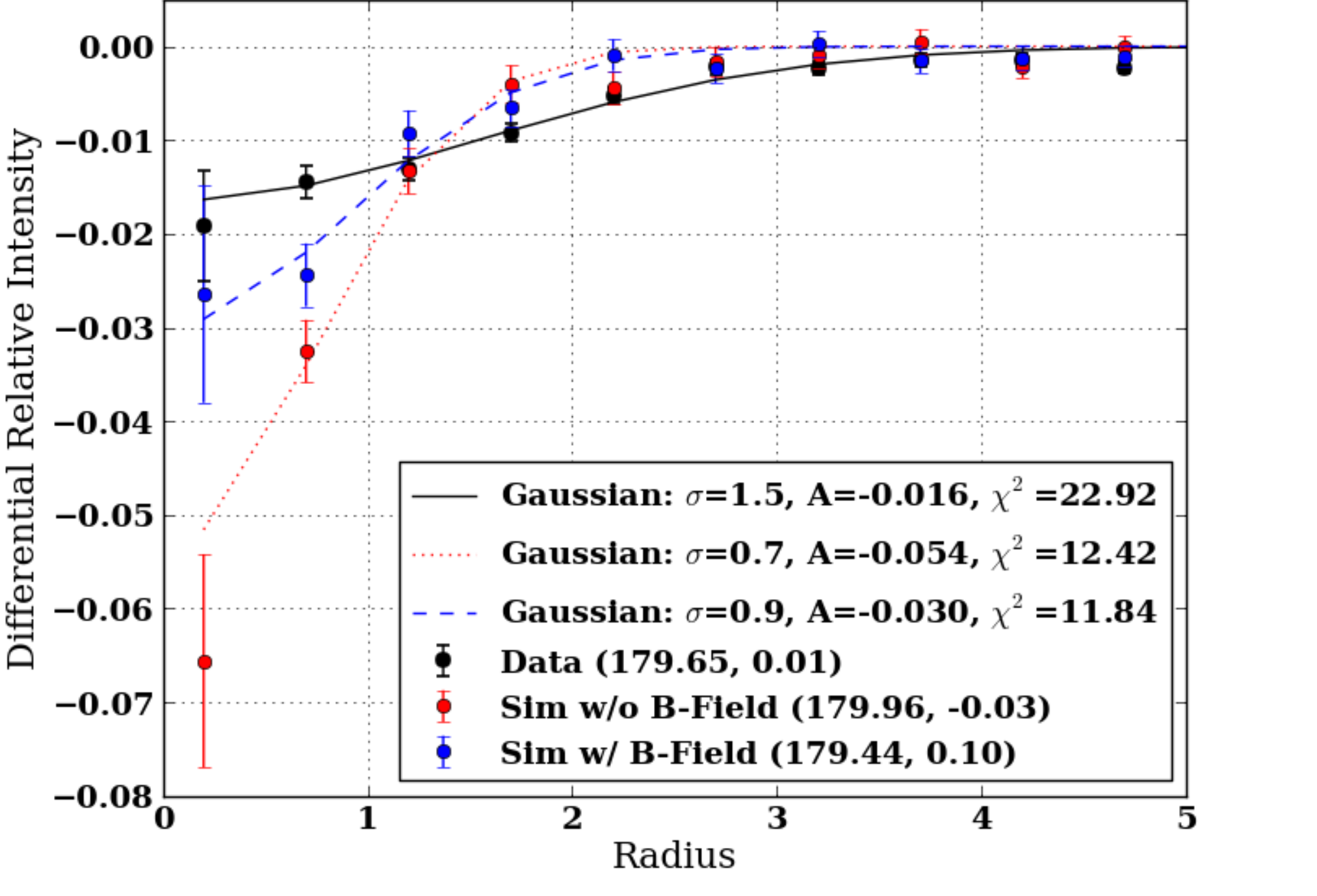}
  \caption{Shown above is the relative intensity as a function of angular distance from the expected Moon shadow position based upon an asymmetric two-dimensional Gaussian fit . A one-dimensional Gaussian fit was applied and the width is reported for the two simulations and the HAWC-30 data. The radial bins are differential so the that the functional form is a simple Gaussian. 
        The simulation widths show the effect of the magnetic field widening the observed Moon shadow width. The disagreement in width between data and simulation is likely due to a mismatch in simulated detector resolution. The relative widening of the simulated moon widths agrees with the ratio of the observed Moon shadow width to the estimated point spread function.}
  \label{fig:geomag_sim_widths}
\end{figure}

The Moon maps for both simulation and data were smoothed in order to find the bin radius for the peak significance. For all maps, this turned out to be a bin radius of 2$^\circ$. For a Gaussian sink, the significance as a function of bin radius follows the formula:
\begin{equation}
 S(r) = \frac{- A\cdot \sigma}{r}(1-e^{\frac{-r^2}{2\sigma^2}}),
\end{equation}
where A is a scaling parameter, $\sigma$ is the width of the sink, and r is the bin radius.

\begin{figure}[t]
  \centering
  \includegraphics[width=0.5\textwidth]{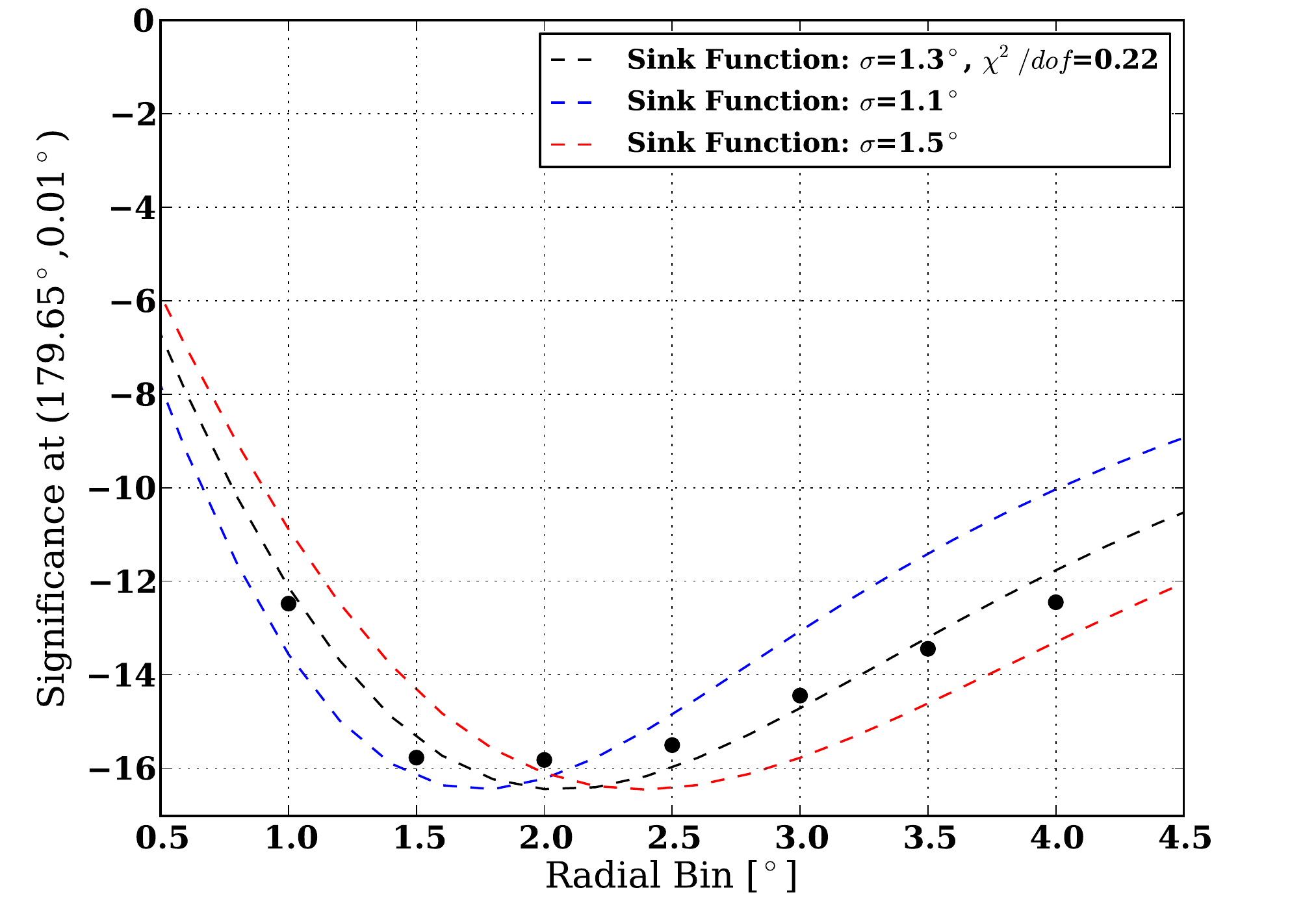}
  \caption{Above is a plot of the Li \& Ma significance at the position of the Moon shadow according to a two-dimensional Gaussian fit for 131 days of HAWC-30 versus the underlying bin radius used for the skymap.}
  \label{fig:smooth_scan}
\end{figure}

A fit to the analytic function for S(r) gives a fit to sigma, the Moon shadow width, of 1.3$^\circ \pm$ 0.2 $^\circ$ (Figure ~\ref{fig:smooth_scan}). This is consistent within errors with the two-dimensional Gaussian fit. The fit is good until r = 4.0$^\circ$, after which the significance of the Gaussian sink fails to drop off quickly enough. This is likely due to the large variance in energies of the dataset.

\section*{Conclusions}

The detection of the cosmic-ray Moon shadow with HAWC-30 is consistent with our simulations and reveals a working detector with an angular resolution of about 1.2$^\circ$. Using a rudimentary energy proxy, we are able to select subsets of cosmic rays with varying median energies. With HAWC-100 scheduled to start operation in August 2013, we can expect more data with a well-behaved and well-understood detector that will continue to improve in angular and energy resolution as well as provide gamma-ray discrimination.

\section*{Acknowledgments}
We acknowledge the support from: US National Science Foundation (NSF); US Department of Energy Office of High-Energy Physics; The Laboratory Directed Research and Development (LDRD) program of Los Alamos National Laboratory; Consejo Nacional de Ciencia y Tecnolog\'ia (CONACyT), M\'exico; Red de F\'isica de Altas Energ\'ias, M\'exico; DGAPA-UNAM, M\'exico; The University of Wisconsin Alumni Research Foundation; and Institute of Geophysics, Planetary Physics at Los Alamos National Lab

\clearpage


\newpage
\setcounter{section}{1}
\nosection{Observation of the Anisotropy of Cosmic Rays with HAWC\\
{\footnotesize\sc Segev BenZvi, Daniel Fiorino, Kathryne Sparks}}
\setcounter{section}{0}
\setcounter{figure}{0}
\setcounter{table}{0}
\setcounter{equation}{0}
%
%
%

\title{Observation of the Anisotropy of Cosmic Rays with HAWC}

\shorttitle{HAWC Cosmic Ray Anisotropy}

\authors{
Segev BenZvi$^{1}$,
Daniel Fiorino$^{1}$,
Kathryne Sparks$^{2}$
for the HAWC Collaboration.
}

\afiliations{
$^1$ WIPAC and Department of Physics, University of Wisconsin-Madison, Madison, WI
USA \\
$^2$ Department of Physics, The Pennsylvania State University, University Park, PA USA \\
}

\email{sybenzvi@icecube.wisc.edu}

\abstract{
The High-Altitude Water Cherenkov (HAWC) Gamma-Ray Observatory
is sensitive to the flux and arrival direction distribution of charged cosmic
rays in the TeV energy band. While the observatory is only partially deployed,
with 30 out of 300 water Cherenkov detectors in data acquisition since
September 2012, HAWC is recording air showers from cosmic rays at a rate above
2~kHz. As a result, we have already accumulated one of the largest data sets of
TeV cosmic rays ever produced. We have analyzed the data and observed a
significant anisotropy at the $10^{-3}$ level in the arrival directions of the
cosmic rays on both large scales ($>60^\circ$) and small scales ($<20^\circ$).
We present these results and compare our findings to previous observations
of anisotropy by experiments such as Milagro, Tibet/ARGO, and others in the
northern hemisphere, and the IceCube Neutrino Observatory in the southern
hemisphere.}


\maketitle

\section*{Introduction}

The HAWC detector is currently under construction 4100~m above sea level on the
north slope of Volc\'{a}n Sierra Negra near Puebla, Mexico.  The observatory,
located at $19^\circ$N latitude, is designed to study the sky in gamma rays and
cosmic rays between 50~GeV and 100~TeV.

While cosmic rays are the major source of background in the gamma-ray analysis,
the distribution of the arrival directions of the cosmic rays is itself of
significant interest.  During the past decade a $10^{-3}$ anisotropy in the
arrival direction distribution of the TeV cosmic rays has been measured with
the Tibet AS$\gamma$ array~\cite{Amenomori:2005dy},
Super-Kamiokande~\cite{Guillian:2007}, Milagro~\cite{Abdo:2008kr,Abdo:2008aw},
EAS-TOP~\cite{Aglietta:2009mu}, MINOS~\cite{DeJong:2011}, and
ARGO-YBJ~\cite{DiSciascio:2013} in the northern hemisphere, and in the southern
hemisphere with the IceCube~\cite{Abbasi:2010mf,Abbasi:2011dc,Abbasi:2012} and
IceTop~\cite{Aartsen:2012} detectors.

The anisotropy has been observed on large angular scales ($>60^\circ$) and
small scales ($<20^\circ$) by multiple experiments.  The large-scale structure
is dominated by dipole and quadrupole moments and does not appear to persist
above the TeV band~\cite{Aartsen:2012}.  Although the large-scale structure is
not well understood, it has long been suggested that weak dipole or dipole-like
features should be a consequence of the diffusion of cosmic rays from nearby
sources in the galaxy~\cite{Erlykin:2006,Blasi:2012}.  The small-scale
structure, on the other hand, could be the product of turbulence in the
galactic magnetic field~\cite{Giacinti:2011mz}.

Using data from HAWC recorded between January and April 2013, we have
measured the cosmic-ray anisotropy in the TeV band.  Due to the low latitude of
the HAWC site, these data cover a region of the sky previously unobserved by
experiments operating in the northern and southern hemispheres.  In these
proceedings we present the results of a search for anisotropy on large and
small angular scales, and compare the observed anisotropy with previous
measurements of the northern and southern skies.

\section*{The HAWC Detector}

The HAWC Observatory is a 22,000~m$^2$ array of close-packed water Cherenkov
detectors (WCDs).  Each WCD consists of a cylindrical steel water tank $4.5$~m
in height and $7.3$~m in diameter.  A non-reflective plastic liner inside the
tank contains 188,000~liters of purified water, and four photomultipliers are
attached to the liner on the floor of the tank: one central high-quantum
efficiency Hamamatsu 10'' PMT and three Hamamatsu 8'' PMTs.  The PMTs face
upward to observe the Cherenkov light produced when charged particles from air
showers enter the tank.  The signals from each PMT are transferred via analog
cables to a counting house in the center of the array, where the data are
digitized using custom front-end electronics and CAEN VX1190A 128-channel TDCs.
Between September 2012 and April 2013, the observatory was operated with 30
WCDs in data acquisition (HAWC-30).  When construction is complete, the
observatory will comprise 300 water Cherenkov detectors with 1200
photomultipliers.

Triggers for gamma-ray and cosmic-ray air showers are formed with a simple
multiplicity trigger which requires $\geq10$~PMTs to be above threshold within
a sliding time window of $100$~ns.  The trigger rate in HAWC-30 is
approximately 5~Hz.  The data are reconstructed offline, and with 30
WCDs the angular resolution of the air shower reconstruction is approximately
$1.5^\circ$.  We note that this is about an order of magnitude worse than in
the complete array, but it is sufficient to observe the anisotropy of the
cosmic rays.

The analysis presented in this paper uses the data collected during the
operation of HAWC-30 between January 1, 2013 and April 15, 2013.  Cuts in
zenith angle of $<45^\circ$ and number of hit PMTs $\geq15$ are used to remove
poorly reconstructed events from the data.  During this period the detector
collected $2.2\times10^{10}$ well-reconstructed events and exhibited a livetime
of $95$~days.  Using the detector simulation we estimate that the median
energy of the data set is about $2$~TeV.

\section*{Analysis}

\begin{figure*}[t]
  \centering
    \includegraphics[width=0.8\textwidth]{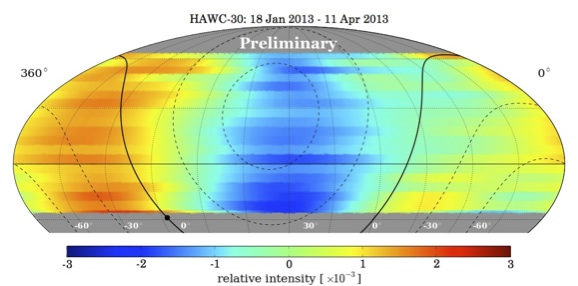}
    \caption{Harmonic fit to the relative intensity of the cosmic rays observed
    in HAWC-30 in 18 independent declination bands.  The sky map is plotted in
    equatorial coordinates; the solid and dashed lines correspond to lines of
    galactic latitude.  The galactic center is shown as a solid black circle.}
    \label{fig:hawc30sky_fb_50day_relint}
\end{figure*}

To produce residual maps of the anisotropy of the arrival directions of the
cosmic rays, we must estimate the expected rate of events in the detector
assuming an isotropic flux.  In order to account for random fluctuations in the
observed rate due to atmospheric effects and the detector geometry, we
calculate the expected flux from the data themselves.  Two analyses -- one
optimized for the measurement of structure on large angular scales, and the
other optimized to observe small scale structure -- have been applied to the
data.  We describe the two analyses and our results in the following sections.

\subsection*{Large-Scale Anisotropy}

For the HAWC-30 data we estimate the large angular scale fractional deviations
from isotropy using the technique of forward-backward
asymmetry~\cite{Abdo:2008aw}.  In this technique we assume that the normalized
intensity $I_\delta$ of the flux of cosmic rays at a fixed declination $\delta$
can be expressed as a three-term harmonic expansion in right ascension $\alpha$:
\begin{equation}
  I_\delta(\alpha) = \frac{R_\delta(\alpha)}{\langle R_\delta(\alpha)\rangle}
  = 1 + \sum_{n=1}^3\gamma_{n,\delta}\cos{n(\alpha-\varphi_{n,\delta})}.
\end{equation}
To estimate the harmonic coefficients $\gamma_{n,\delta}$ and calculate the
residual intensity, we divide the arrival directions of the data along the
local meridian into positive and negative hour angles $\pm\xi$.  For a fixed
time interval characterized by the local sidereal time $\theta_0$, we can
define the relative asymmetry of the arrival directions in the ``forward'' and
``backward'' directions (along and against the rotation of the Earth) by
\begin{equation}
  FB_\delta(\theta_0,\xi) =
    \frac{N_{\theta_0,\delta}(+\xi)-N_{\theta_0,\delta}(-\xi)}
         {N_{\theta_0,\delta}(+\xi)+N_{\theta_0,\delta}(-\xi)}.
\end{equation}
Since $\alpha=\theta_0\pm\xi$ and the residual coefficients $\gamma\ll1$, the
asymmetry can be expressed as
\begin{equation}\label{eq:fb_expansion}
  FB_\delta(\theta_0,\xi)\approx-\sum_{n=1}^3
    \gamma_{n,\delta}\sin{n\xi}\sin{(n(\theta_0-\varphi_{n,\delta}))}.
\end{equation}
In practice, we use the data to produce a two-dimensional table of $FB$ as a
function $\alpha$ and $\xi$ for a fixed declination $\delta$.  Fitting
eq.~\eqref{eq:fb_expansion} to the table provides the fit coefficients
$\gamma_{n,\delta}$ for the different declination bands.

This procedure has been applied to the HAWC-30 data and the result is shown in
Figure~\ref{fig:hawc30sky_fb_50day_relint}.  In our analysis we performed the
fits in 18 independent declination bands.  The fits were performed on data
taken when the detector had at least 102 active PMTs, covering a subset of the
period between January 1 and April 15, 2013.


While the data are independent for each band in $\delta$, we note that all
bands exhibit a relative deficit of events in the interval
$120^\circ<\alpha<240^\circ$ and a relative excess outside this region.  The
fit results closely match those of Milagro, which were produced using the same
technique \cite{Abdo:2008aw}.  In addition, the amplitude of the fit in each
band is approximately $10^{-3}$, which is in agreement with the scale of the
large dipole and quadrupole anisotropies reported in the northern and southern
hemispheres by other experiments.

\subsection*{Small-Scale Anisotropy}

\begin{figure*}[t]
  \centering
  \includegraphics[width=0.8\textwidth]{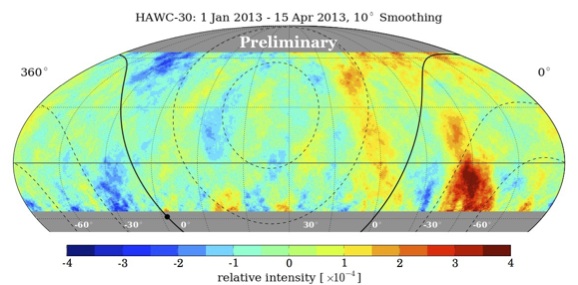}
  \includegraphics[width=0.8\textwidth]{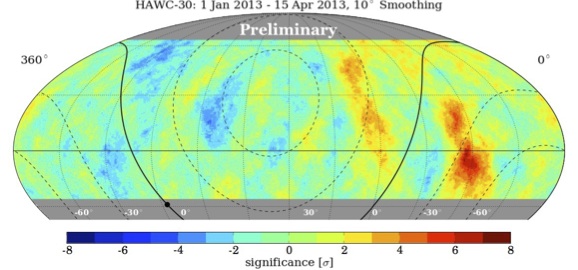}
  \caption{
    {\slshape Top}: relative intensity of the HAWC-30 cosmic-ray data produced
    using a time-integration window $\Delta t=2$~hr.
    {\slshape Bottom}: sky map of the pre-trial significance of the features
    shown in the top panel.
  }
  \label{fig:hawc30sky_2hr_N512_S10}
\end{figure*}

To search for anisotropy on small angular scales, we directly compute the
relative intensity as a function of equatorial coordinates $(\alpha,\delta)$.
We begin by binning the sky into an equal-area grid with a resolution of
$0.1^\circ$ per bin using the HEALPix library \cite{Gorski:2005}.  A binned
data map $N(\alpha,\delta)$ is used to store the arrival directions of air
showers recorded by the detector, and a binned reference map $\langle
N(\alpha,\delta)\rangle$ is computed to describe the arrival direction
distribution if the data arrived isotropically at Earth.

The reference map is produced using the direct integration technique described
in \cite{Atkins:2003}, adapted for the HEALPix grid.  In brief, we proceed by
collecting all events recorded during a predefined time period $\Delta t$ and
integrate the local arrival direction distribution against the detector event
rate.   The method effectively smooths out the true arrival direction
distribution in right ascension on angular scales of roughly $\Delta
t\cdot15^\circ~\text{hr}^{-1}$ such that the analysis is only sensitive to
structures smaller than this characteristic angular scale.  The direct
integration procedure also compensates for variations in the detector rate
while preserving the event distribution in declination.  

Once the reference map is obtained, we calculate the deviations from isotropy
by computing the relative intensity
\begin{equation}
  \delta I_i(\alpha,\delta)
    = \frac{\Delta N_i}{\langle N\rangle_i}
    = \frac{N_i(\alpha,\delta) - \langle N_i(\alpha,\delta)\rangle}
           {\langle N_i(\alpha,\delta)\rangle},
\end{equation}
which gives the amplitude of deviations from the isotropic expectation in each
angular bin $i$.  The significance of the deviation can be calculated using
the method of Li and Ma~\cite{LiMa:1983}.

The analysis was carried out on HAWC-30 data using $\Delta t=2~\text{hr}$ to
obtain sensitivity to features smaller than $30^\circ$ in right ascension.  The
results are plotted in Figure~\ref{fig:hawc30sky_2hr_N512_S10}, with the
relative intensity shown on the top and the significance shown on the bottom.
The data have been smoothed using a $10^\circ$ top-hat function to make the
clustering of arrival directions readily apparent.

Several prominent features are visible in the sky map, notably the regions of
excess flux at $\alpha=60^\circ$ and $\alpha=120^\circ$.
The number of independent pixels in the sky map is of order $10^5$, and after
accounting for trial factors only these two regions of excess are
significant at $>5\sigma$.  These hot spots correspond to the
$10^\circ$-$20^\circ$ regions of cosmic-ray access observed by Milagro (Regions
A and B in \cite{Abdo:2008kr}) and ARGO-YBJ~\cite{DiSciascio:2013}.

An inspection of the sky maps in Figure~\ref{fig:hawc30sky_2hr_N512_S10} also
shows that every region of excess is associated with a neighboring deficit in
the cosmic ray flux.  If these small-scale features are produced by turbulence
in the galactic magnetic field, as proposed in \cite{Giacinti:2011mz}, then
there is no reason for us to favor hot spots over cold spots in the analysis.
However, we do find that none of the deficit regions is currently more
significant than $-5\sigma$ after trial factors are taken into account.  In
addition, the estimation of reference maps with direct integration can be
biased by a strong anisotropy, leading to artificial deficits or excesses next
to regions of true excess or deficit \cite{Abdo:2008kr}.  For this reason we
cannot currently rule out the possibility that the deficit regions are
artifacts of the analysis rather than real features in the residual cosmic ray
flux.

\section*{Discussion}

While the HAWC-30 data are still not highly significant when compared to
published observations of the cosmic ray anisotropy, it is still instructive to
compare our results (particularly the small-scale structure) with other
measurements in the northern and southern hemispheres.

When comparing Fig.~\ref{fig:hawc30sky_2hr_N512_S10} to published results from
the northern hemisphere, two features stand out.  The first is the weak hot
spot at $\alpha=250^\circ$.  This region of excess does not appear in the
10~TeV Milagro data, but the ARGO-YBJ collaboration has observed a hot spot in
the same part of the sky~\cite{DiSciascio:2013}.  Future data from HAWC will
confirm whether the excess is a real effect or a fluctuation.

The second feature of interest is the elongated excess at $\alpha=120^\circ$,
which is visible at all declination angles observed by HAWC.  The extension of
the excess to high northern declinations is notable because this is not
observed in the Milagro data, even though Milagro was located at $35^\circ$N
latitude.  However, the significance maps published by ARGO-YBJ
also extend to high declinations, and these maps indicate that the Region B
excess extends to the northern edge of the ARGO exposure.

With the present data it is not clear if these differences and similarities are
significant.  However, if we take them at face value then there are several
possible explanations for the features we observe.  HAWC and ARGO-YBJ have a
similar energy threshold (which is lower than that of Milagro), and both
detectors are at similar geomagnetic latitudes.  Therefore if there is a
significant contribution to the anisotropy from events at the energy threshold
it could explain the similarity between observations at the two sites.  We will
investigate this possibility with future data.

A comparison between data from HAWC-30 and IceCube is very interesting because
the HAWC data cover declinations down to $-25^\circ$, which is the edge of the
exposure region in the published IceCube results.  When comparing the data sets
we find that the gross features of the cosmic ray sky maps, such as a
significant large-scale deficit near $\alpha=180^\circ$ and the presence of
$10^\circ$-$20^\circ$ structures, are visible in both hemispheres.  In
particular it is interesting to observe that the elongated Region B excess at
$\alpha=120^\circ$ extends into the southern hemisphere with a similar relative
intensity as that observed in the northern sky.

There are some differences between the data sets.  Both the large and
small-scale structures appear to be misaligned in $\alpha$, with the structures
observed in IceCube appearing at higher right ascensions by
$\sim15^\circ$-$20^\circ$.  In addition, the Region A excess, which is the most
prominent feature in the HAWC-30 sky maps, does not appear to have a
counterpart in the IceCube data.

The source of these differences is unclear but there are several possible
explanations.  First, the IceCube data are of relatively high energy with
respect to other measurements of the anisotropy.  The median energy of the
low-energy IceCube cosmic-ray analysis is 20~TeV, as compared to 2~TeV for
HAWC-30.  As a result the differences in the sky maps could be due to the
different energy scales of the data sets.  As HAWC accumulates large statistics
it should be possible to check this assumption by producing a high-energy sky
map.

A second source of difference could be a composition/trigger bias between the
two detectors.  For example, the IceCube detector observes cosmic-ray air
showers via the TeV muons which travel $1.4$~km into the south polar ice sheet.
Hence, at the energy threshold of the cosmic-ray analysis the IceCube detector
will preferentially trigger on proton events over heavier nuclei (which produce
lower-energy muons on average) \cite{Abbasi:2011dc}.  Recent data indicate that
the TeV band is a complex region of changing cosmic-ray composition, with
protons becoming the sub-dominant primary type at energies above
10~TeV~\cite{Ahn:2010gv,Adriani:2011cu}.  As a result, IceCube and HAWC may not
be observing a completely equivalent population of cosmic rays.

\section*{Conclusions}

Using the 30-tank configuration of HAWC we have observed a significant
large-scale and small-scale anisotropy in the arrival direction distribution of
the cosmic rays in the TeV band.  Our observations are largely in agreement
with previous measurements of the anisotropy in the northern and southern
hemispheres.  The areas of disagreement, such as the possibility of a third
region of significant small-scale excess and discrepancies between HAWC and
IceCube, may be due to the presence of unaccounted energy and composition
effects in the anisotropy.  Both possibilities will be the subject of a future
detailed investigation.

\vspace*{0.5cm}

\clearpage


\newpage
\setcounter{section}{2}
\nosection{HAWC and Solar Energetic Transient Events\\
{\footnotesize\sc Alejandro Lara}}
\setcounter{section}{0}
\setcounter{figure}{0}
\setcounter{table}{0}
\setcounter{equation}{0}
%
%
%
%
\title{HAWC and Solar Energetic Transient Events}

\shorttitle{SEPs and HAWC}

\authors{
Alejandro Lara$^{1}$,
for the HAWC Collaboration.
}

\afiliations{
$^1$ Institute of Geophysics, National University of Mexico, UNAM \\
}

\email{alara@geofisica.unam.mx}

\abstract{
The High Altitude Water Cherenkov (HAWC) observatory is being constructed 
at the volcano Sierra Negra (4100 m a.s.l.) in Mexico. HAWC's primary purpose 
is the study of both galactic and extra-galactic sources of high energy gamma 
rays. The HAWC instrument will consist of 300 large water Cherenkov detectors 
whose counting rate will be sensitive to cosmic rays with energies above the 
geomagnetic cutoff of the site ($\sim 8$ GV). In particular, HAWC will 
detect solar energetic particles known as Ground Level Enhancements (GLEs), 
and the effect of Coronal Mass Ejections on the galactic cosmic rays, 
known as Forbush Decreases (FDs). The Milagro experiment, the HAWC predecessor, 
successfully observed GLEs and the HAWC engineering array "VAMOS" already 
observed a FD. HAWC will be sensitive to  $\gamma$ rays and 
neutrons produced during large solar flares. 
In this paper, we present the instrument and discuss 
its capability to observe 
solar energetic events. i. e., flares and CMEs.}


\maketitle

\section*{Introduction}
The most energetic transient events in the solar system: flares and coronal 
mass ejections (CMEs), accelerate particles up to tens of GeV. 
In the current scenario, electric fields associated to the magnetic reconnection
during flares accelerate particles up to hundreds of MeV 
\cite{bib:aschwanden12}. 
These particles may interact with ambient nuclei in the solar atmosphere 
and generate both $\gamma$ rays and neutrons which, in turn may escape  
from the solar atmosphere and the most energetic ones can be detected 
at ground level by air showers  
arrays such as HAWC.

Some of these particles may  escape to the interplanetary medium 
and may interact with an interplanetary CME  (ICME). 
Both: stochastic acceleration in the turbulent ICME plasma and 
Fermi acceleration in the shock driven by the ICME are plausible 
 \cite{bib:bombardieri06,bib:bombardieri07,bib:bombardieri08},  
generating in this way the so called solar 
energetic particles (SEPs), which travel along the interplanetary magnetic 
field lines. 


If these magnetic field lines are connected to the Earth, the SEPs will 
enter the magnetosphere and will produce air showers in the atmosphere. 
Finally these secondary particles will increase the counting rate in the 
cosmic-ray detectors on Earth, this is known as 
ground level enhancements (GLEs).

On the other hand, CMEs are huge amounts of plasma and magnetic field 
traveling through the interplanetary medium, swiping away and/or trapping 
galactic cosmic rays (GCRs). This is observed as a decrease in the measured 
GCR flux in the inner heliosphere, called Forbush decreases (FDs).

Among other effects of solar activity, 
GLEs and FD have been observed for decades using principally Neutron Monitors 
(NMs). It is important to note that a worldwide network of 
NMs is necessary to measure the characteristics of these events - energy 
spectrum, temporal and spatial shape. 

In this sense a detector which is able to give information about 
the GLE anisotropy
and  energy spectrum with  high time resolution is necessary.
In this paper we present and discuss the capabilities of the new 
``High Altitude 
Water Cherenkov'' (HAWC) gamma ray detector to observe and study GLEs and FDs.

\section*{HAWC}
As the name states HAWC is being constructed at 4100 m above the sea level 
on the ``Sierra Negra'' Volcano in the South-East  central part of Mexico 
[$18^\circ 59' 41'' N, 97^\circ 18' 28''W$].
HAWC will consist of 300 water Cherenkov detectors (WCD). Each one is a 
cylindrical container of 4.5 m height and 7.3 m diameter filled with 
purified water and equipped with four photo-multipliers (PMTs). 
HAWC will cover an area of 22 000 $m^2$. A detailed description of HAWC is 
presented in \cite{bib:Mostafa}.

HAWC has two modes of data acquisition, a TDC and scaler. In TDC mode an event 
is defined and stored when certain number of PMTs are hit by the same 
cascade (within a certain time window).
In the scaler mode the rate of each PMT is counted and stored with a time 
resolution of few milliseconds.

The scaler system stores separately the signal of each PMT giving the 
opportunity of configuring different sizes or effective areas of 
the detector.
The scaler will be an important system for solar observations. This 
system is  sensitive to secondary muons created by GCR and SEPs. 

We started test observations with a tenth of the array, i.e. 30 WCD, 
which is called HAWC30. The observations presented in this work were 
made with HAWC30.

\subsection*{Solar Mode}

The primary goal of HAWC is the detection of high energy $\gamma$-rays.  
The TDC system 
is optimized for events with energy higher than $\sim$ 
200 GeV. For solar studies there will be a buffer where the raw data will be 
stored for one or two days and the threshold will be adjusted  to 
be sensitive to lower energies ($\> 8$ GV which is the geomagnetic cutoff 
of the site).

%
If solar activity is 
reported (a large solar flare with associated SEPs or a Forbush Decrease), 
the data will be analyzed using lower thresholds, to allow the detection of 
low energy particles.

\section*{Forbush Decreases}
\begin{figure}[t]
  \centering
  \includegraphics[width=0.45\textwidth]{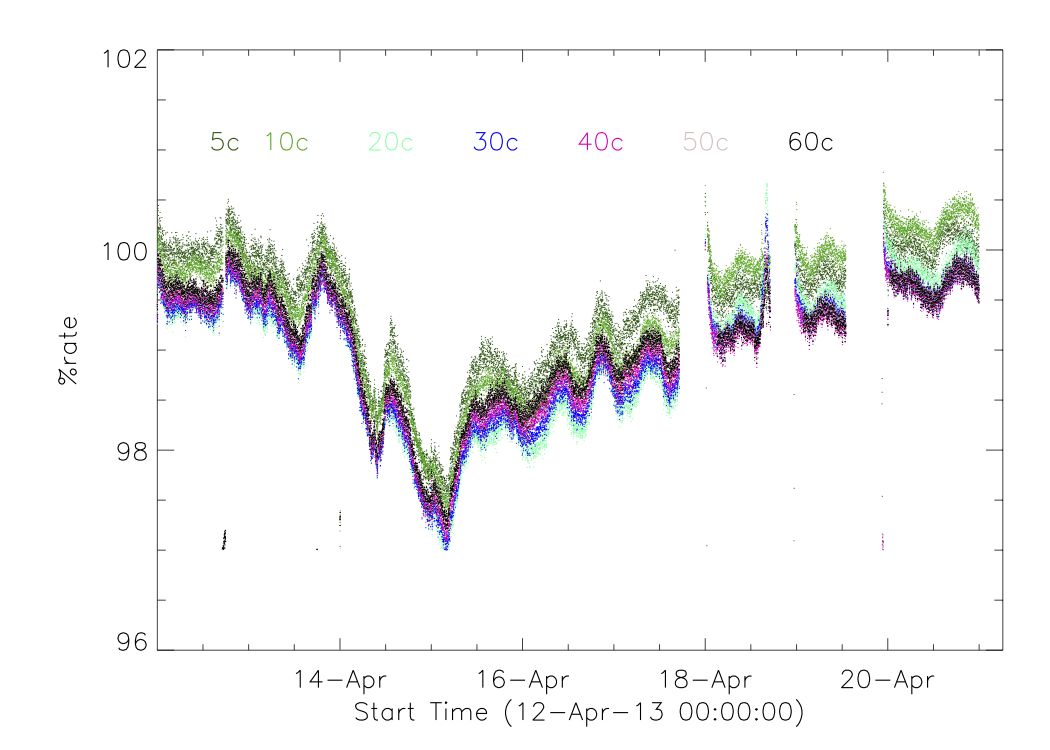}
  \caption{Forbush Decrease observed by different sub-arrays of  HAWC30.}
  \label{fig:fd1}
 \end{figure}

HAWC will be a large air shower detector well situated to study  solar activity.  
In particular FD will be observed and studied. 
As an example of the capabilities of HAWC, in Figure \ref{fig:fd1} we present 
a FD observed by HAWC30. 

In this example we have constructed seven 
sets, 
containing the 5, 10, 20, 
30, 40, 50 60 PMTs (or channels) shown in different colors in 
Figure  \ref{fig:fd1}.

The sub-second rate of each individual channel were 
integrated to obtain a time resolution of one minute and  then, each channel 
was corrected for barometric pressure and temperature variations. 
Finally,  we constructed the sets by computing the average of all channels 
in the set every minute.

The elements of each set were selected randomly but each minor set is
a subset of the major one. Once we have a larger array we have to be 
more careful by selecting the sets of PMTs or sub-arrays, since these 
sub-arrays will give us information about the energy spectrum of the GCR 
modulated by  solar activity.

The  time and energy resolution of HAWC will be an excellent  tool to
observe and study the FD precursors, that is, the particles trapped by 
the ICME magnetic field and then ejected  probably due to  a loss cone 
effect \cite{bib:asipenka2009}.

\begin{figure}[t]
  \centering
  \includegraphics[width=0.45\textwidth]{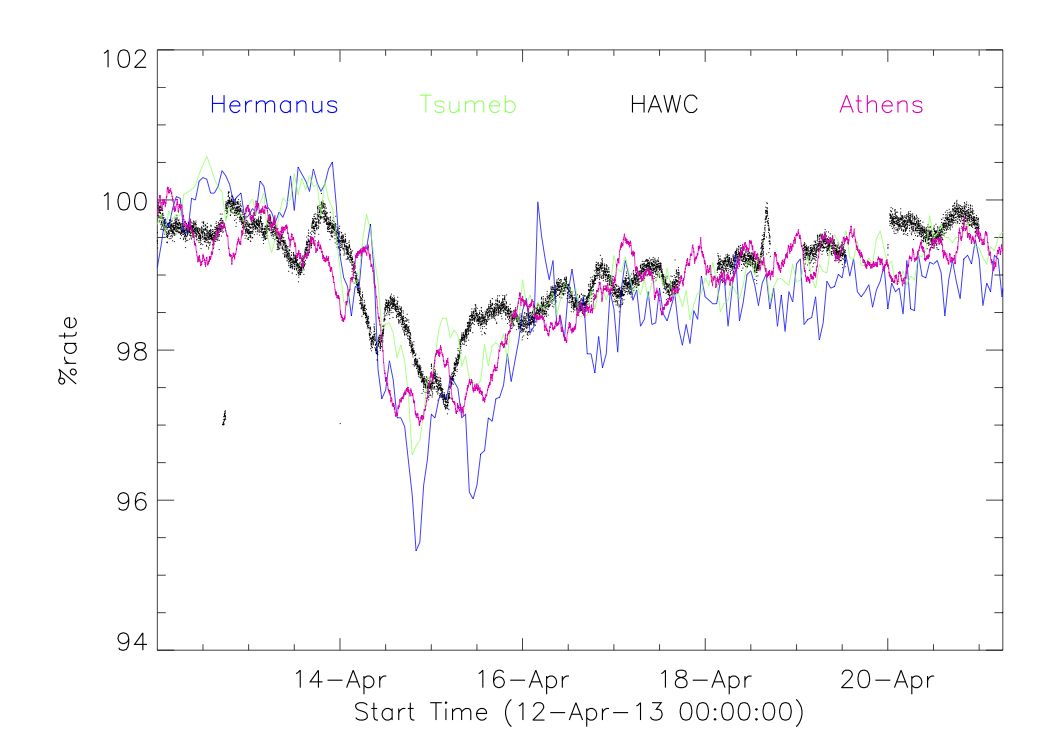}
  \caption{Forbush Decrease observed by HAWC30 and three different NMs:
Athens, Hermanus and Tsumeb with 8.5, 4.5 and 8.9 cutoff rigidity, 
respectively.}
  \label{fig:comp}
 \end{figure}

Figure \ref{fig:comp} shows the rate (percentage) observed by three neutron 
monitors
\footnote{Hermanus and Tsumeb  data come from http://www.nwu.ac.za/neutron-monitor-data, whereas Athens data come from http://www.nmdb.eu.}
and HAWC 60-PMT rate. The cutoff 
rigidity of both Athens and Tsumeb NMs are similar to the HAWC one. 
Therefore the time profile and 
maximum decrease are similar.

\section*{Ground Level Enhancements}

The energy resolution of HAWC,  and its better response  to higher 
energies (compared to neutron monitors), will allow us to address one 
of the major outstanding questions of SEPs, that is, the upper energy 
limit of the solar eruptive events. 

Milagro, the  predecessor of HAWC, observed the 2005 
January 20 GLE; and the analysis performed using different PMT 
multiplicity \cite{bib:morgan2008} 
shows the potential of this kind of arrays to provide information on the 
primary GLE energy spectrum.
 
\begin{figure}[t]
  \centering
  \includegraphics[width=0.45\textwidth]{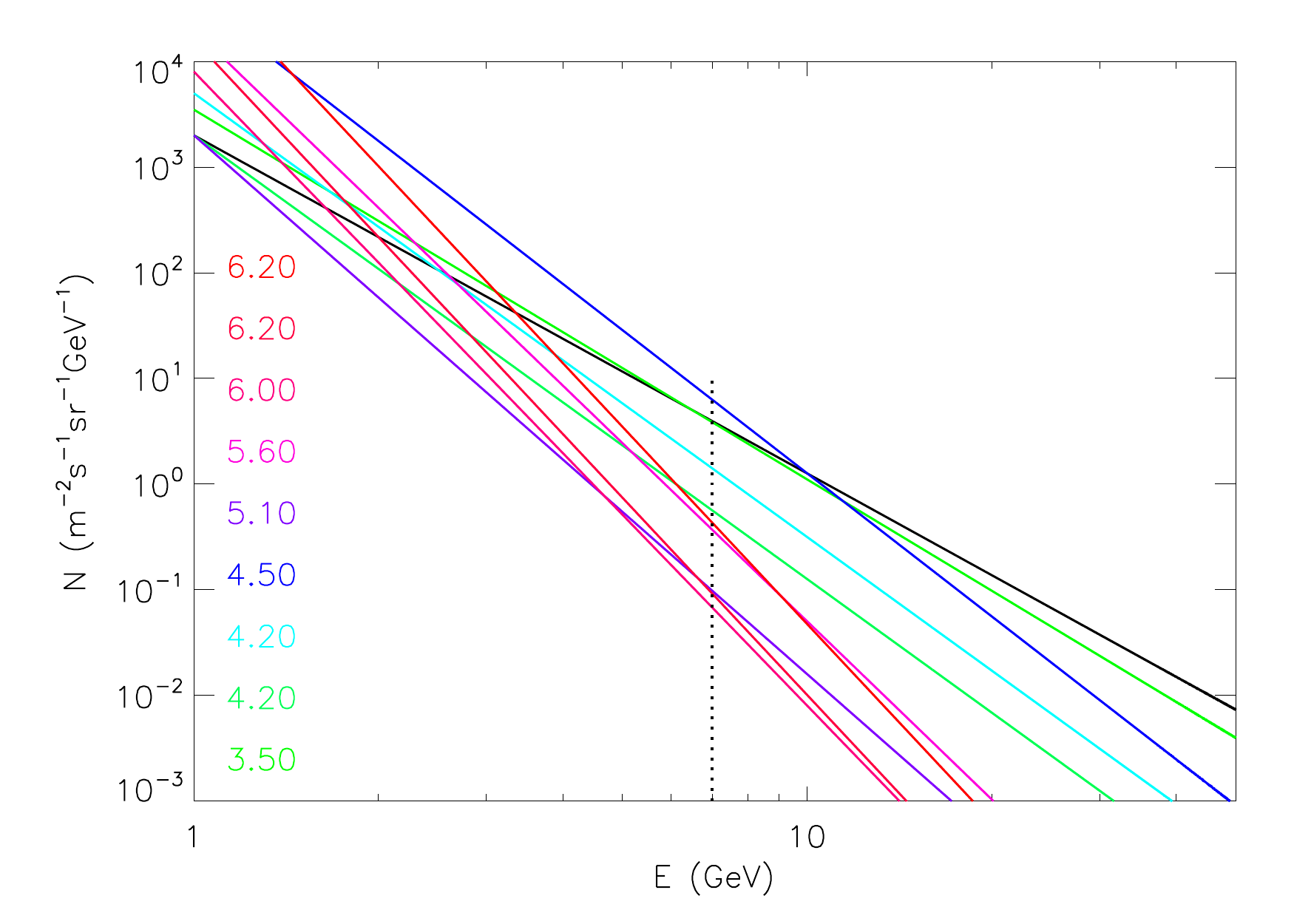}
  \caption{High energy spectrum of eleven GLEs reported by 
\cite{bib:Vashenyuk} and assuming that the high energy cut off is above 50 GeV. 
The vertical line mark the geomagnetic cut off energy for protons at the HAWC 
site.}
  \label{fig:spectra}
 \end{figure}

The solar spectrum is softer than the GCR spectrum as shown in Figure 
\ref{fig:spectra} where we have plotted the power law ($J = J_1E^{-\gamma}$) 
 fitted to 
11 GLEs reported by \cite{bib:Vashenyuk}. Here $J_1$ is a normalization 
constant and $\gamma$ is the spectral index which is marked with different 
color in Figure \ref{fig:spectra}. 
%
%
%
%
It can be seen that the energy spectra of GLEs extend beyond the 
geomagnetic cutoff, into the energy range accessible to HAWC.

If we know the energy spectrum  of the primaries, we can determine the 
degree of  anisotropy of the GLE, as discussed in the next section where 
we present the asymptotic cones of acceptance of HAWC.

\subsection*{Asymptotic directions at HAWC site} 
\begin{figure}[t]
  \centering
  \includegraphics[width=0.4\textwidth]{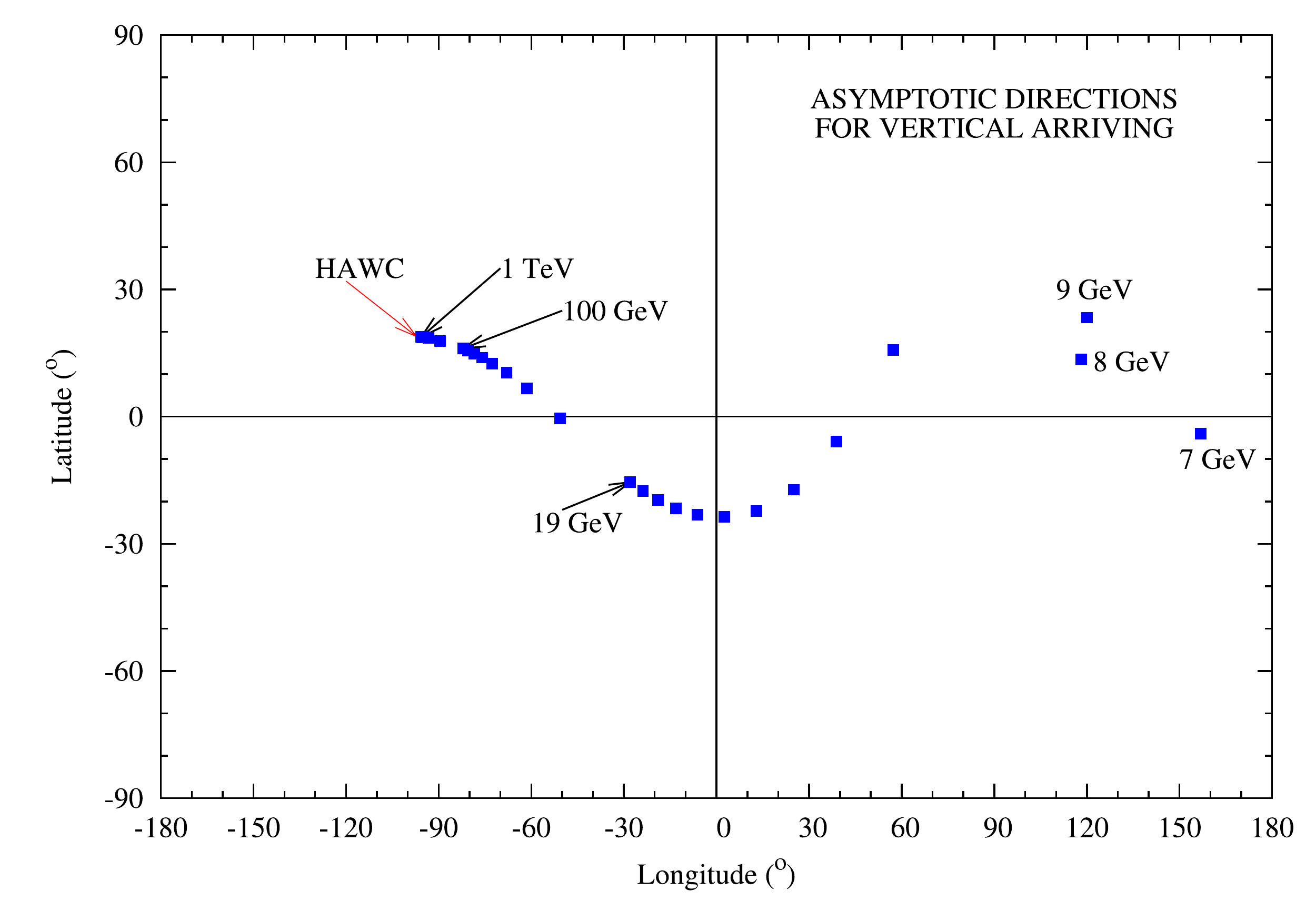}
  \caption{Asymptotic cones for protons arriving at HAWC location.}
  \label{fig:dirasint}
 \end{figure}

We computed the asymptotic cones for protons arriving at HAWC location 
(vertically incident at 20 km a.s.l.), using the geomagnetic reference 
field of 1995. The results appear in Figure \ref{fig:dirasint}. 
As we can expect, very high energy protons (above 1 TeV) practically 
reach HAWC with no deflection in the Earth magnetic field,  
while very low energy ones are deflected by a large angle. 

\begin{figure}[t]
  \centering
  \includegraphics[width=0.4\textwidth]{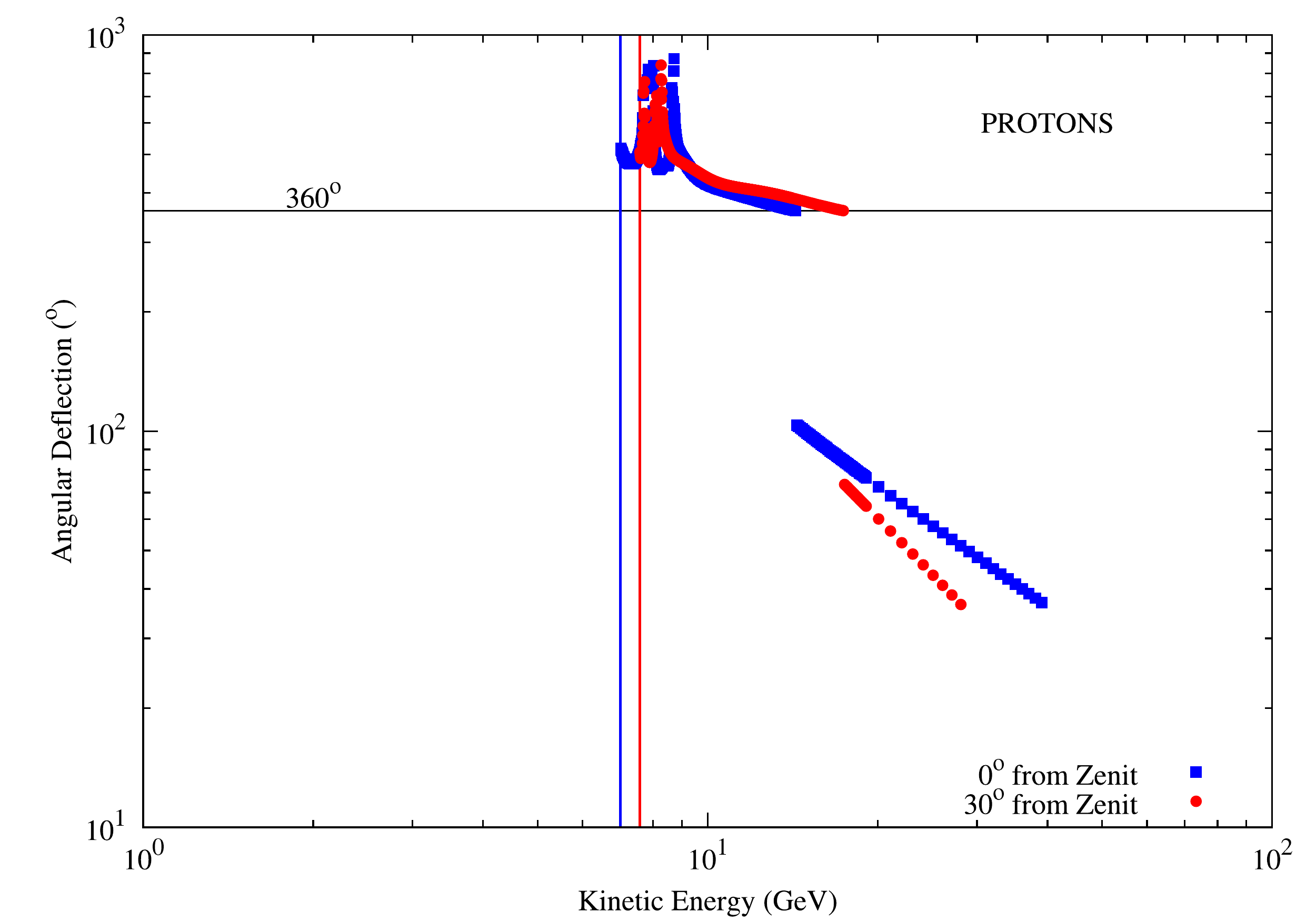}
  \caption{The central angular deflection.}
  \label{fig:desvang}
 \end{figure}

The cutoff energy for vertically incident protons at HAWC is about 7 GeV 
(rigidity = 7.9 GV). Protons with energy below 15 GeV reach HAWC after a 
complete orbit around the Earth. This can be seen in Figure \ref{fig:desvang}, 
where we plotted the central angular deflection as a function of proton 
kinetic energy for two arrival directions at HAWC, 1) vertically incident 
protons (in blue) and 2) protons arriving at angle of $30^\circ$ from the vertical 
direction (in red). 
The central angular deflection is the angle between the arrival direction at 
HAWC location and the proton asymptotic direction at the top of the 
magnetosphere (about 25 Earth radii). 
As we can see the deflection angle increases with decreasing energy. 
For protons arriving at $30^\circ$ from the vertical direction, the cutoff 
energy is 7.6 GeV (rigidity = 8.5 GV, shown by vertical red line).

\section*{Solar $\gamma$-rays and Neutrons}
 
During large solar flares, ions with energy  in excess of 1 GeV are produced.
These ions collide with the ambient nuclei generating  neutrons 
and $\gamma$-rays. 
%
%
%

In particular $\gamma$-rays above 100 MeV where observed by  SMM, Compton 
and now by Fermi.
These photons are produced through the decay of charged and neutral pions, 
which, in turn, have been produced by much higher energy ions. 
Charged pions decay into
electrons that radiate, while neutral pions decay directly into photons, 
usually heavily Doppler broadened. 

Our ability to deduce the nature of parent ion population
responsible for the $\gamma$ rays is limited by the confounding 
multiple processes that
separate the ion population from the consequent photons. 
However, when neutrons
are produced, which should be every time pions are produced, we have
complementary information about the ion spectrum. The energy of the neutron is
more closely tied to that of the parent ion and the angular distribution of 
neutrons
should differ significantly from that of the pion-based photons. The two
measurements together can be used to deduce the pitch angle distribution and 
other
factors that manifest themselves in anisotropic emission.

We propose to use  HAWC, 
as well as low-latitude and preferably high altitude neutron monitors, 
to register a
solar neutron signal. 


\section*{Summary}
The High Altitude Water Cherenkov Array will start operating by the 
end of 2013, 
giving us the opportunity of observing high energy solar transients 
during the declining phase of solar cycle 24.
Then HAWC will operate for 5 years up to the rising part of solar cycle 25.

In this way we expect to have the opportunity of observe several 
GLEs and many more 
FDs with high energy, time and angular resolution. This is a great 
opportunity to address outstanding questions about particle acceleration by 
solar flares and coronal mass ejections and about the large 
scale magnetic structures which cause the GCR solar modulation.

Because of its altitude, large effective area and low latitude, HAWC is well 
suited to detect solar energetic 
events. 
Combined with ground level and space borne particle detectors 
HAWC will provide the most constraining data on the 
very high
energy ion population in flares and shed more light on the question of 
whether this
high-energy emission is a result of back-diffused protons from a CME shock.
%
A detailed study of HAWC and neutron monitors response will be 
performed and combined with solar charge particles, neutrons and 
$\gamma$ ray models to form a complete picture of the processes occuring.



\section*{Acknowledgment}

We thank CONACyT 179588 and DGAPA IN112412 for partial support. The ICRC 2013
is funded by FAPERJ, CNPq, FAPESP, CAPES and IUPAP.

\clearpage


\newpage
\setcounter{section}{3}
\nosection{Observation of the March 2012 Forbush Decrease with the Engineering
Array of the High-Altitude Water Cherenkov Observatory\\
{\footnotesize\sc Mario Castillo, Humberto Salazar, Luis Villase\~{n}or}}
\setcounter{section}{0}
\setcounter{figure}{0}
\setcounter{table}{0}
\setcounter{equation}{0}
%
%

\title{Observation of the March 2012 Forbush decrease with the engineering array of the High Altitude Water Cherenkov Observatory}

\shorttitle{Forbush decrease with the HAWC engineering array}

\authors{
M. Castillo$^{1}$,
H. Salazar$^{1}$,
L. Villase\~nor$^{2}$
for the HAWC Collaboration$^{3}$.
}

\afiliations{
$^1$ Facultad de Ciencias F\'isico Matem\'aticas, BUAP, Av. San Claudio y 18 Sur, Colonia San Manuel, Ciudad Universitaria, Puebla, M\'{e}xico\\
$^2$ Instituto de F\'isica y Matem\'aticas, UMSNH, Edificio C-3, Ciudad Universitaria, Morelia, M\'{e}xico\\
$^3$ For a complete author list, see the special section of these proceedings
\scriptsize{
}
}

\email{mcastillomaldonado@yahoo.com}

\abstract{Neutron monitors have reported the observation of a Forbush decrease on March 7, 2012.
VAMOS, an engineering array built for the HAWC (High Altitude Water Cherenkov) Observatory, was operational at that time.
This array was composed of six water Cherenkov detectors located at the HAWC site near the volcano Sierra Negra in Mexico.
VAMOS had two data acquisition (DAQ) systems, one designed to readout full air shower events (Main DAQ) and the other designed to monitor the count rates of the individual PMTs (Scaler DAQ).
We have analyzed data from both the Main DAQ and the Scaler DAQ systems.
We present a comparison between the observation of this transient event in VAMOS and neutron monitor detectors located in Mexico City and the South Pole.
We also describe the necessary corrections in the count rates due to atmospheric effects.}


\maketitle

\section*{Introduction}
Decreases in the cosmic-ray count rate which last typically for about one week, were first observed by Forbush (1973) \cite{Forbush37} and Hess and Demmelmair \cite{Hess37} using ionization chambers.
There are two basic types of Forbush decreases. 'Non-recurrent decreases' are caused by transient interplanetary events (shock and ejecta) which are related to mass ejections from the Sun. They have a sudden onset,
reach a maximum depression within about a day and have a more gradual recovery. 'Recurrent decreases' have a more gradual onset, are more symmetric in profile,
and are well associated with co-rotating high speed solar wind streams.
The amplitude of the daily variation for cosmic rays observed in neutron detectors generally increases during the recovery phase of a Forbush decrease due to the presence of the anisotropy
caused by the transient interplanetary structure propagating beyond the Earth's orbit, which produces a decrease of cosmic ray flux arriving from that direction \cite{hilary}.
In the next sections we explain how to correct the data from the two DAQ systems from the VAMOS array by atmospheric pressure in order to observe a Forbush decrease that occurred on March 7, 2012 in coincidence with the neutron detector from the Cosmic Ray Observatory located at UNAM and the McMurdo station at the South Pole.
\section*{The High Altitude Water Cherenkov Observatory} 

The HAWC observatory is a facility designed to observe TeV gamma-rays and cosmic-rays with an instantaneous aperture that covers more than 15\% of the sky.
With this large field of view, the detector will be exposed to half of the sky during a 24-hour period.
HAWC is under construction by a collaboration of scientists from Mexico and USA at Sierra La Negra volcano near Puebla, Mexico, at an altitude of 4100 m a.s.l.
When completed, HAWC will consist of 300 water Cherenkov detectors (WCDs) of 7.3 m in diameter by 4.5 m depth with a light tight bladder and 3 peripheral and 1 central photomultiplier tubes (PMTs) facing upwards from the bottom each.
It will survey the sky in search of steady and transient gamma-ray sources in the 0.1-100 TeV energy range. 

\begin{figure}[!t]
  \centering
  \includegraphics[width=0.4\textwidth]{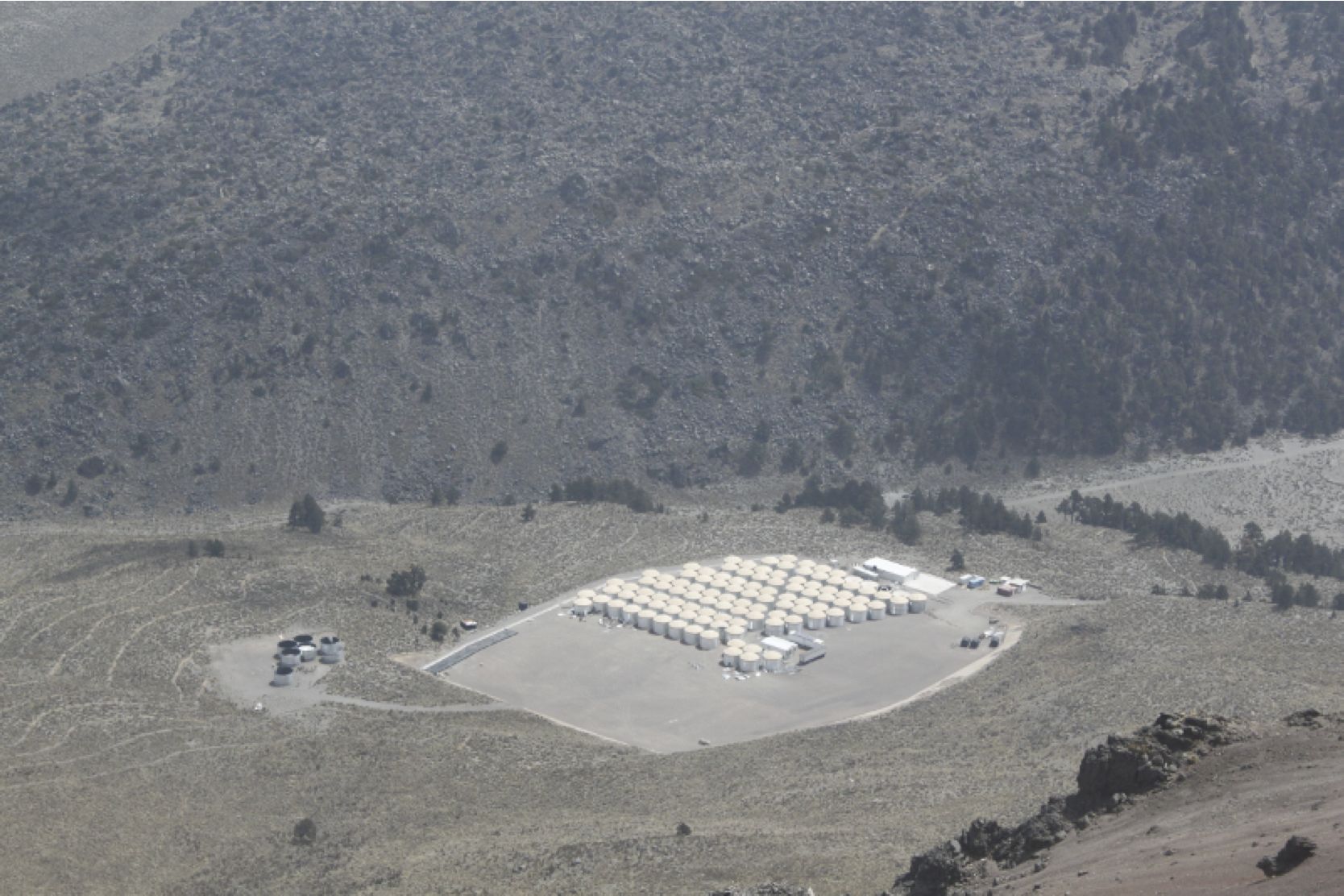}
  \caption{The HAWC observatory site at 4100 m a.s.l. This picture shows 106 tanks constructed by May 16, 2013. It is also showed the 6-tanks engineering array VAMOS at the left of the picture.}
  \label{hawc_s}
 \end{figure}
\subsection*{The VAMOS array}
An engineering array of six tanks, called VAMOS (Verification And Measuring of Observatory Systems), was built on site.
Six of the tanks were filled with filtered water and instrumented with 4 to 7 PMTs per tank. Engineering data have been collected with 6 tanks.
Continuous operation of VAMOS started in Sept 29th, 2011 and finished operation in March 2012. 
The VAMOS array was installed aside the HAWC site. See Figure ~\ref{hawc_s}.
\subsection*{Operation principle}
\begin{figure}[!t]
  \centering
  \includegraphics[width=0.4\textwidth]{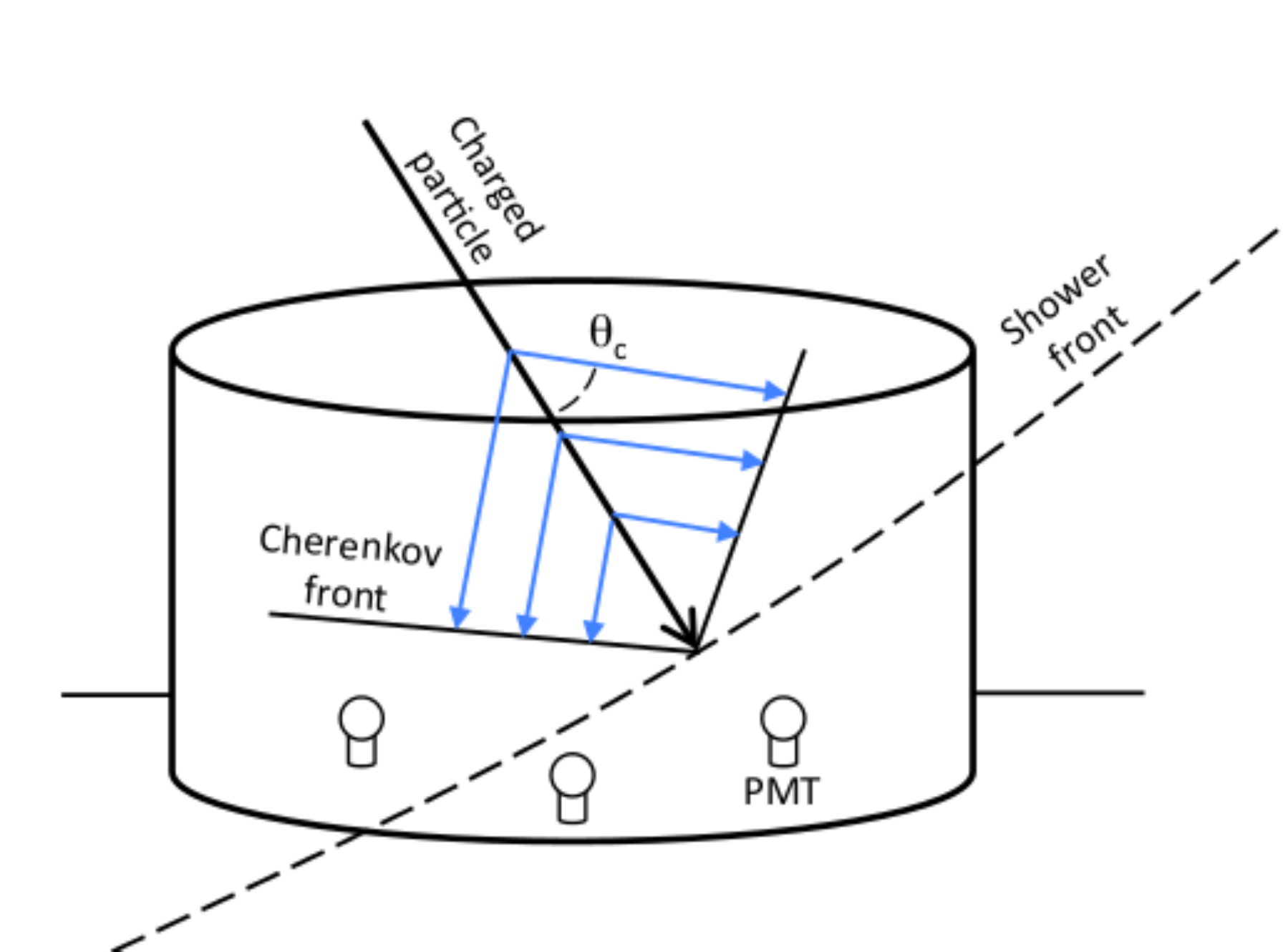}
  \caption{The Cherenkov effect in Water Cherenkov Detectors.}
  \label{operation_p}
 \end{figure}

Relativistic charged particles from extended air showers originated by primary cosmic-rays produce Cherenkov radiation as the air shower cross the WCDs.
Cherenkov radiation is emitted at a precise angle $\theta_{c}$ with respect to the particle trajectories and it is detected by photomultiplier tubes at the bottom of the tank. See Figure ~\ref{operation_p}.

\subsection*{The Main DAQ system}
HAWCs primary DAQ system records individual events produced by air showers which are large enough to simultaneously illuminate a significant fraction of the HAWC array.
In the simplest approach, depending on the number of hit PMTs during a given time window (trigger condition), a trigger will be issued and sent to time-to-digital converters (TDCs). The TDCs store the measured times of the PMT pulses occurring close to the trigger time. The data of each issued trigger are called an event.
The event data recorded will consist of time stamps of the leading and trailing edges from the discriminated PMT pulses. The PMT pulses are discriminated in 2 thresholds; the lowest one set to 1/4 photon-electron (resulting in 2-edges hits) and the highest one set to 5 photon-electrons (resulting in 4-edges hits).

\subsection*{The Scaler DAQ system}
During regular operations, the counts in HAWCs PMTs are due to cosmic-ray air showers, naturally occurring radioactivity near the PMTs and thermal noise in the PMTs.
By monitoring PMTs at the lowest threshold (\textgreater 1/4 photoelectron), this system can be used to cross-check the proper functioning of the detector and to monitoring the secondary cosmic-rays background.
The solar activity must be detectable by measuring this background. All PMTs pulses occurring above the lowest theshold are counted in 10 ms time windows.

\section*{Count rate corrections} 
\begin{figure}[t]
  \centering
  \includegraphics[width=0.4\textwidth]{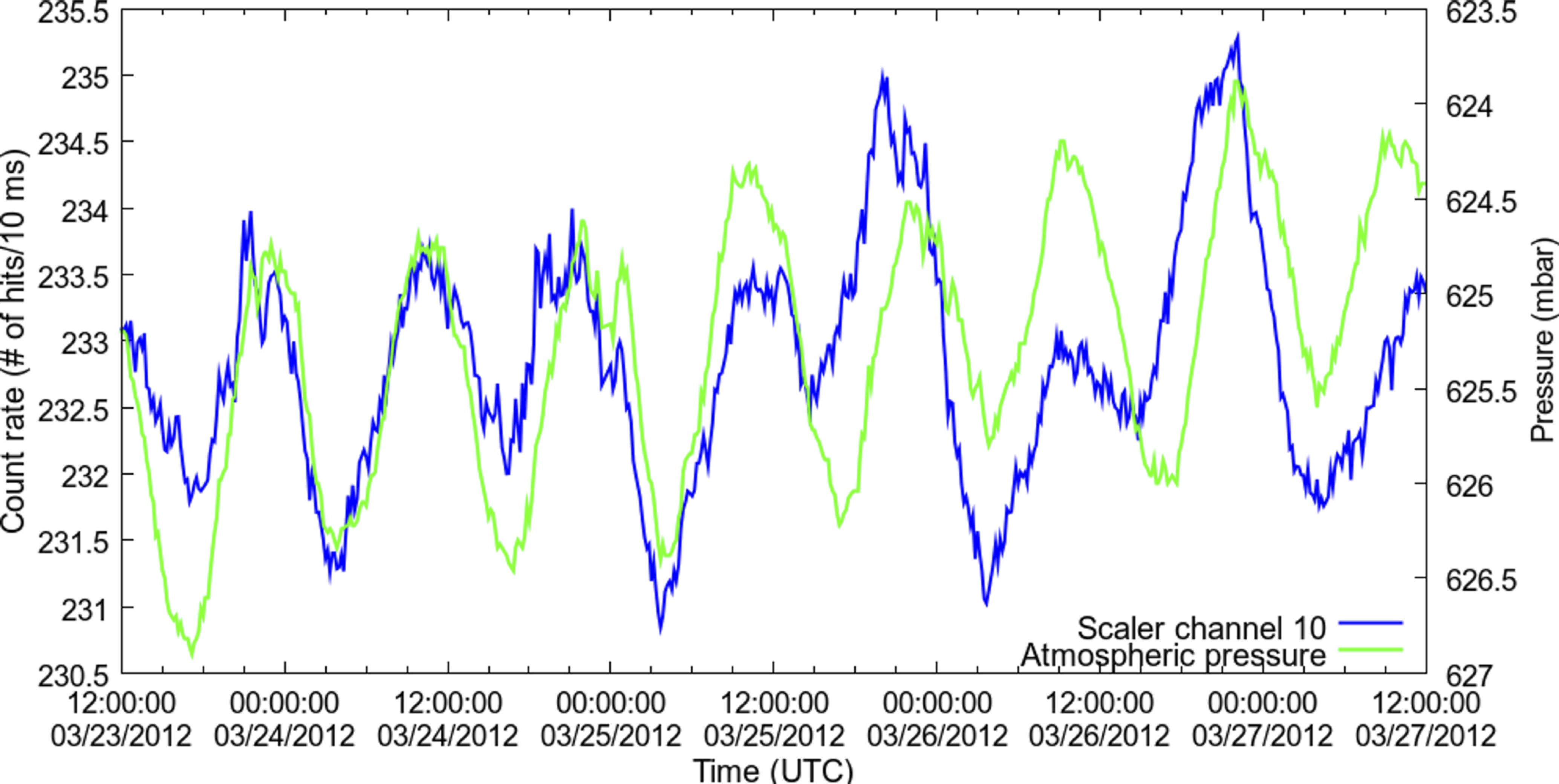}
  \caption{Relationship between the count rate for the Scaler DAQ channel 10 and the atmospheric pressure.}
  \label{scaler_r}
 \end{figure}

\begin{figure}[t]
  \centering
  \includegraphics[width=0.4\textwidth]{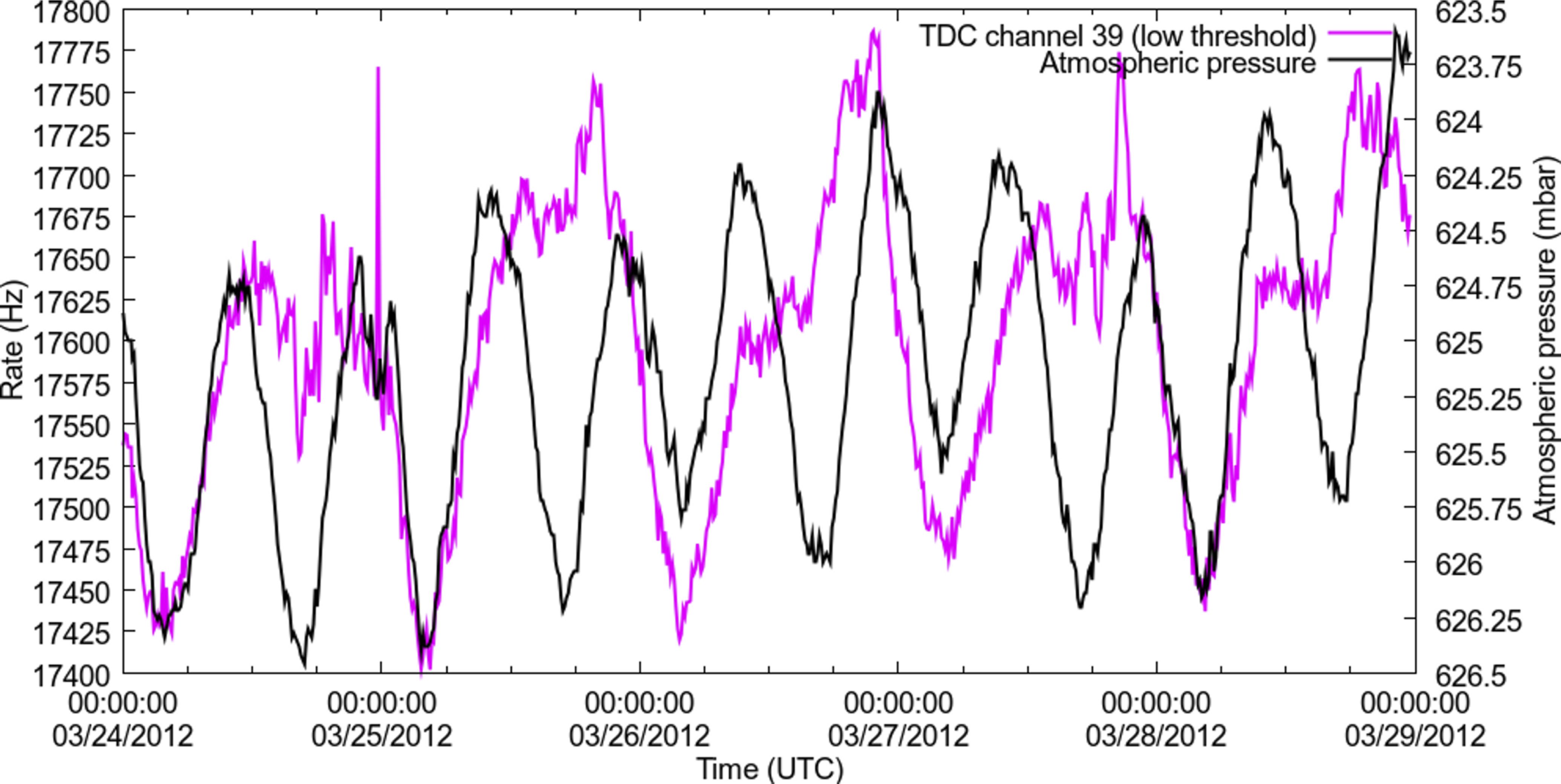}
  \caption{Relationship between the count rate for the Main DAQ channel 39 and the atmospheric pressure. Here the count rate is the number of 2-edge hits per second.}
  \label{tdc_r2}
 \end{figure}

\begin{figure}[t]
  \centering
  \includegraphics[width=0.4\textwidth]{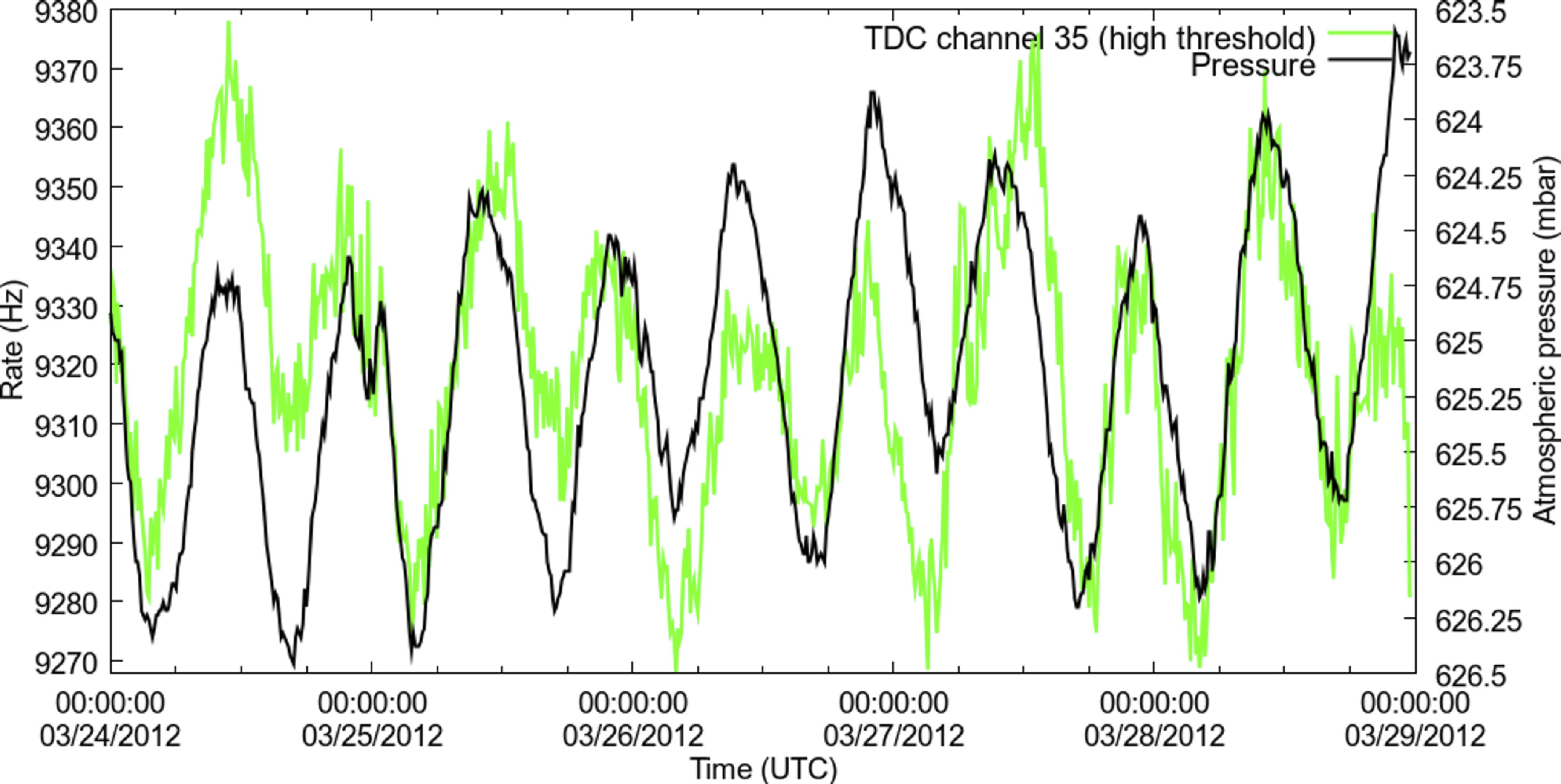}
  \caption{Relationship between the count rate for the Main DAQ channel 35 and the atmospheric pressure. Here the count rate is the number of 4-edge hits per second.}
  \label{tdc_r4}
 \end{figure}

This section describes the correction process done on the count rates (Main DAQ and Scaler DAQ systems) from March, 2012.
We focus on the detection of the Forbush decrease that started on March 7, 2012 and lasted about two weeks.
The first step is to deduce the relationship between the count rate and the variations of atmospheric pressure.
To do that we have chosen a period where the data were stable (not affected by the Forbush decrease). This period is from March 24, 2012 to March 28, 2012.
We then averaged the raw data from a system of sensors that record the atmospheric pressure and temperature, Scaler and Main DAQ systems in periods of 10 minutes.
It is worth to mention that quality cuts were applied to the data from the system of sensors and the DAQ systems.
Examples for this relationship are the Figures ~\ref{scaler_r} for Scaler DAQ system and ~\ref{tdc_r2} and ~\ref{tdc_r4} for Main DAQ system, where there is an anti-correlation between the count rate and the atmospheric pressure. Note that the atmospheric pressure axis was inverted to make this anti-correlation more evident.

\begin{figure}[t]
  \centering
  \includegraphics[width=0.4\textwidth]{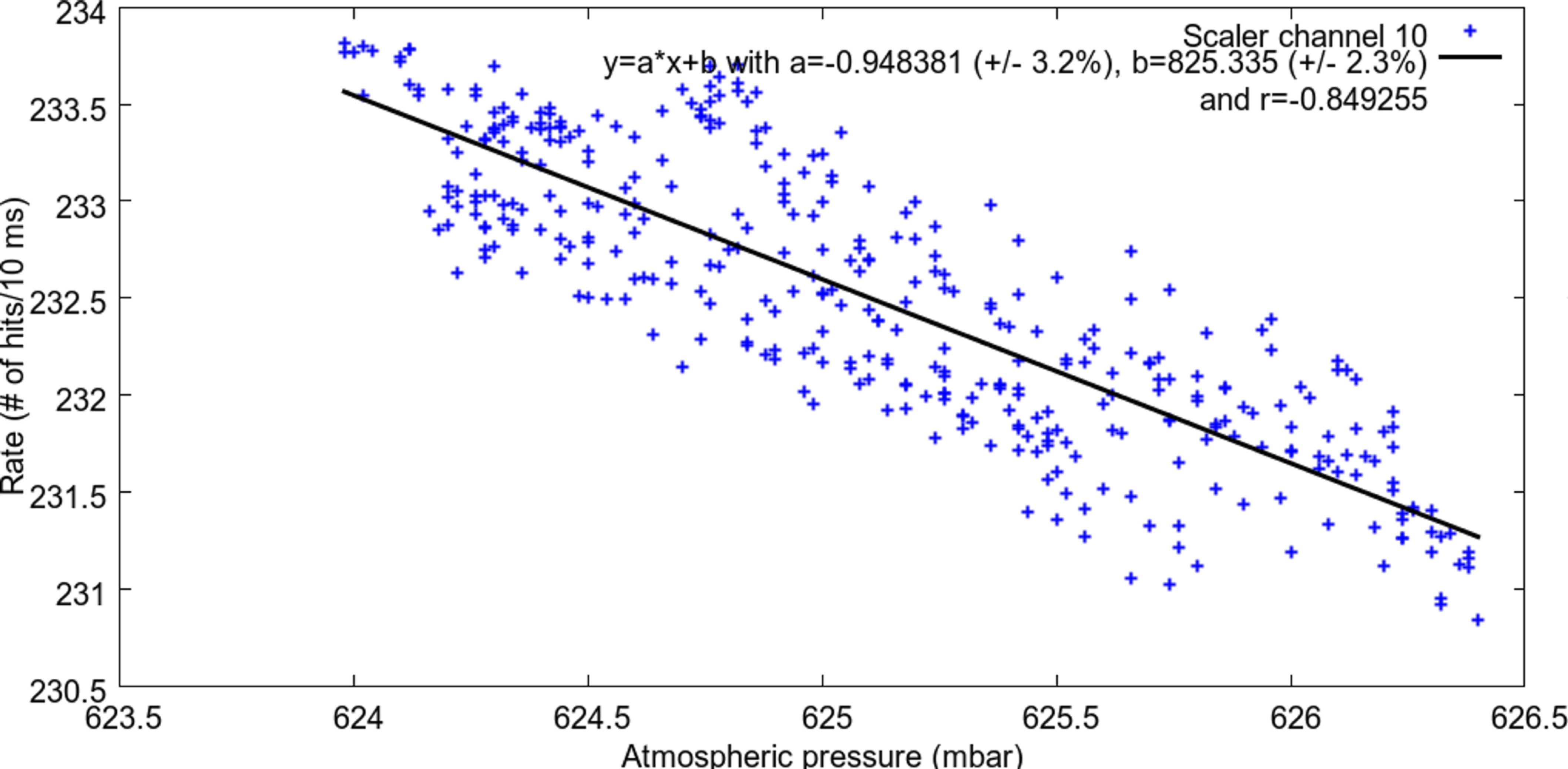}
  \caption{Count rate vs atmospheric pressure for the Scaler DAQ channel 10. Both, the linear correlation coefficient (r $\sim$ -0.849255) and the fact that the two  variables fit in a line with a negative slope indicate a strong, inverse, linear relationship between both observables.}
  \label{scaler_pressure}
 \end{figure}

\begin{figure}[t]
  \centering
  \includegraphics[width=0.4\textwidth]{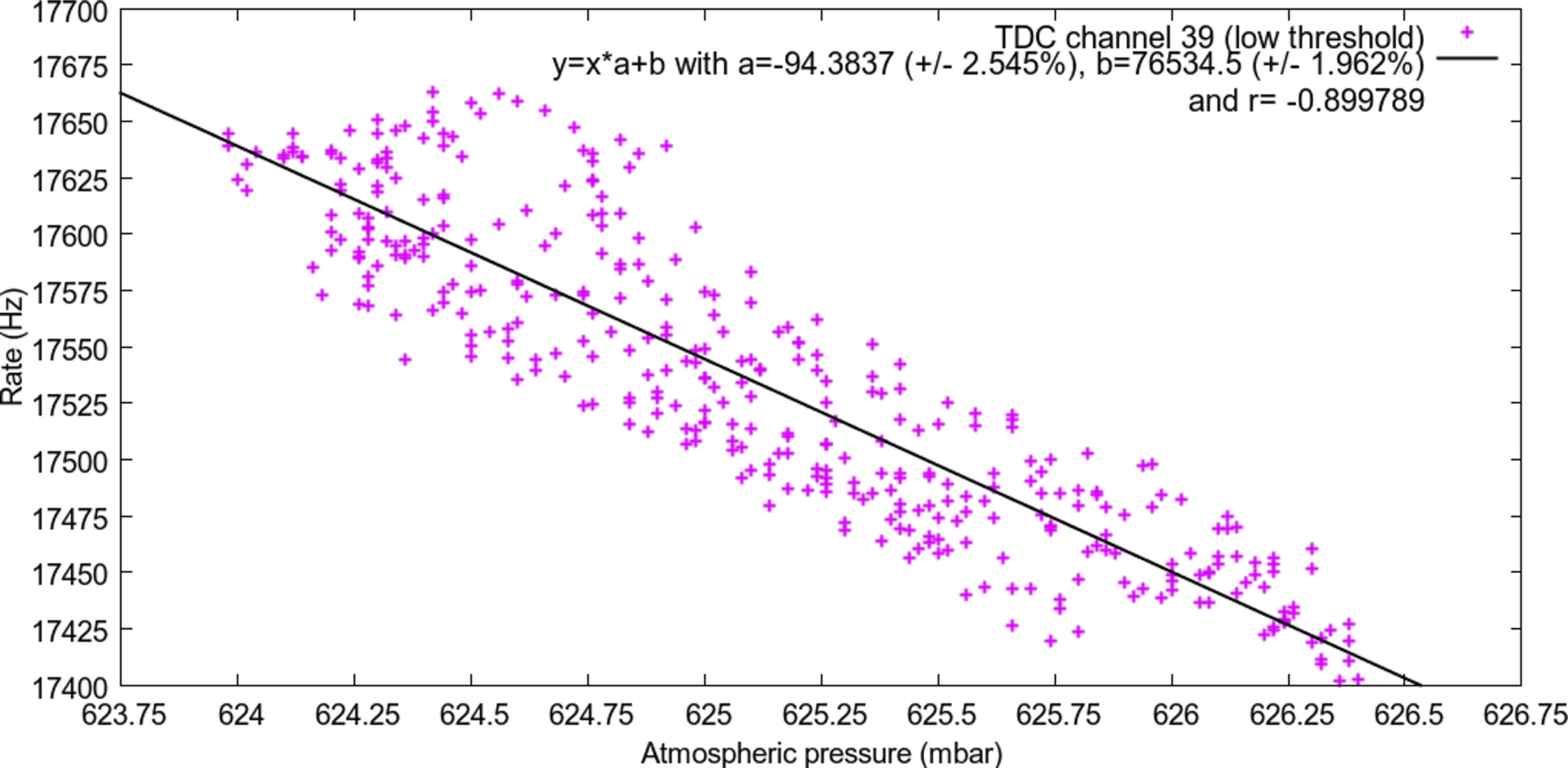}
  \caption{Count rate vs atmospheric pressure for the Main DAQ channel 39. Here the count rate is the number of 2-edge hits per second. Both, the linear correlation coefficient (r $\sim$ -0.899789) and the fact that the two  variables fit in a line with a negative slope indicate a strong, inverse, linear relationship between both observables.}
  \label{tdc_pressure2}
 \end{figure}

\begin{figure}[t]
  \centering
  \includegraphics[width=0.4\textwidth]{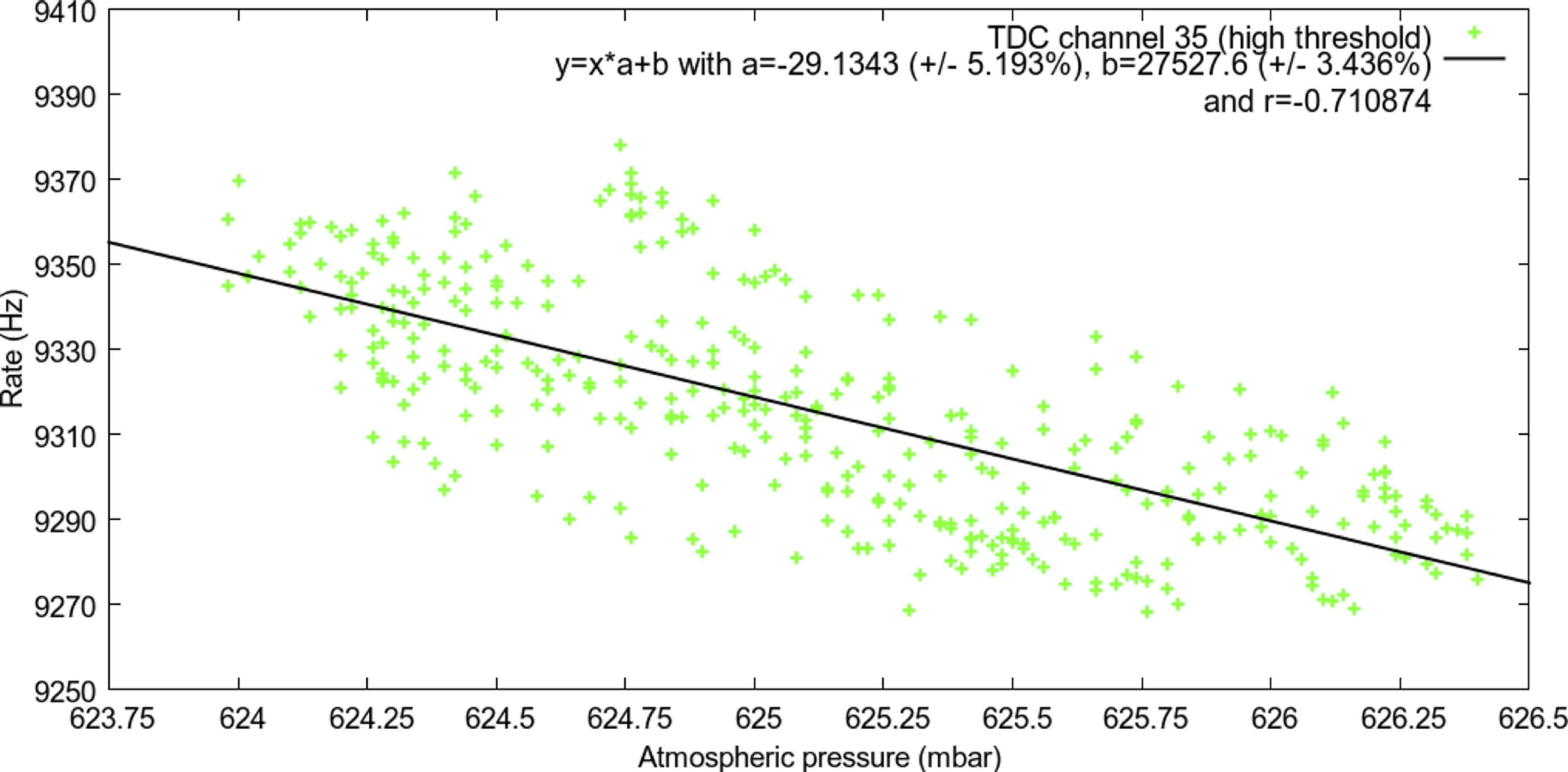}
  \caption{Count rate vs atmospheric pressure for the Main DAQ channel 35. Here the count rate is the number of 4-edges hits per second. Both, the linear correlation coefficient (r $\sim$ -0.710874) and the fact that the two  variables fit in a line with a negative slope indicate a strong, inverse, linear relationship between both observables.}
  \label{tdc_pressure4}
 \end{figure}

After consider three sub-periods of these data (daytime, nighttime, all day time) it turned out that the relation between the count rate and the atmospheric pressure is linear and inverse, and the best sub-period to see this was nighttime.
This relationship is showed at Figures ~\ref{scaler_pressure} and ~\ref{tdc_pressure2} and ~\ref{tdc_pressure4} for the Scaler and the Main DAQ systems respectively.

A general explanation of this relationship is the next: The atmospheric pressure indicates the quantity of matter (air) is above the detector.
When the atmospheric pressure increases, the quantity of matter passed by the secondary particles also increases, resulting that a larger quantity of these particles been absorbed.
When the atmospheric pressure decreases, the quantity of matter passed by the secondary particles also decreases, resulting in more secondary particles get the detector.

Knowing that the relation between these variables is linear and anti-correlated we proceed to correct the count rate by pressure.
Then, the second step is to calculate the parameters (slope and constant) for the line that fits in the count rate and the pressure plot for the nighttime data.
This relationship is characterized by the Formula 1, where the parameters a and b were measured for each channel.
\begin{equation}
   R_{P}= a*P+b 
\end{equation}
with
\begin{eqnarray*}
R_{P} & = & \textrm{count rate due to atmospheric pressure} \\
      & = & [hits/\textrm{10 ms}] \\
    a & = & \textrm{fit slope} \\
      & = & {[hits/\textrm{10 ms}][g/cm^{2}]}^{-1} \\
    P & = & \textrm{atmospheric pressure} \\
      & = & [g/cm^{2}] \\
    b & = & \textrm{fit constant} \\
      & = & [hits/\textrm{10 ms}]
\end{eqnarray*}
The errors on the fit slope and intercept parameters are indicative of the quality of the procedure to correct the count rates to remove variations due to atmospheric pressure. A total of 26 PMTs for the Scaler DAQ system and 20 PMTs for the Main DAQ system showed errors in this step lower than 7 \%. The remaining PMTs were either off or did not registered any rate variation.
The third step is to correct the data for the whole month by applying Formula 2.
\begin{equation}
R_{corr}= R_{uncorr}-R_{P}+<R_{uncorr}>
\end{equation}
with
\begin{eqnarray*}
R_{corr} & = & \textrm{Count rate corrected by pressure} \\
 & = & [hits/\textrm{10 ms}] \\
R_{uncorr} & = & \textrm{Count rate not corrected by pressure} \\
 & = & {[hits/\textrm{10 ms}]}
\end{eqnarray*}
This procedure is done channel by channel for both DAQ systems. 
\section*{Comparison between the VAMOS array and neutron monitors}
\begin{figure*}[t]
 \centering
  \includegraphics[width=0.57\textwidth]{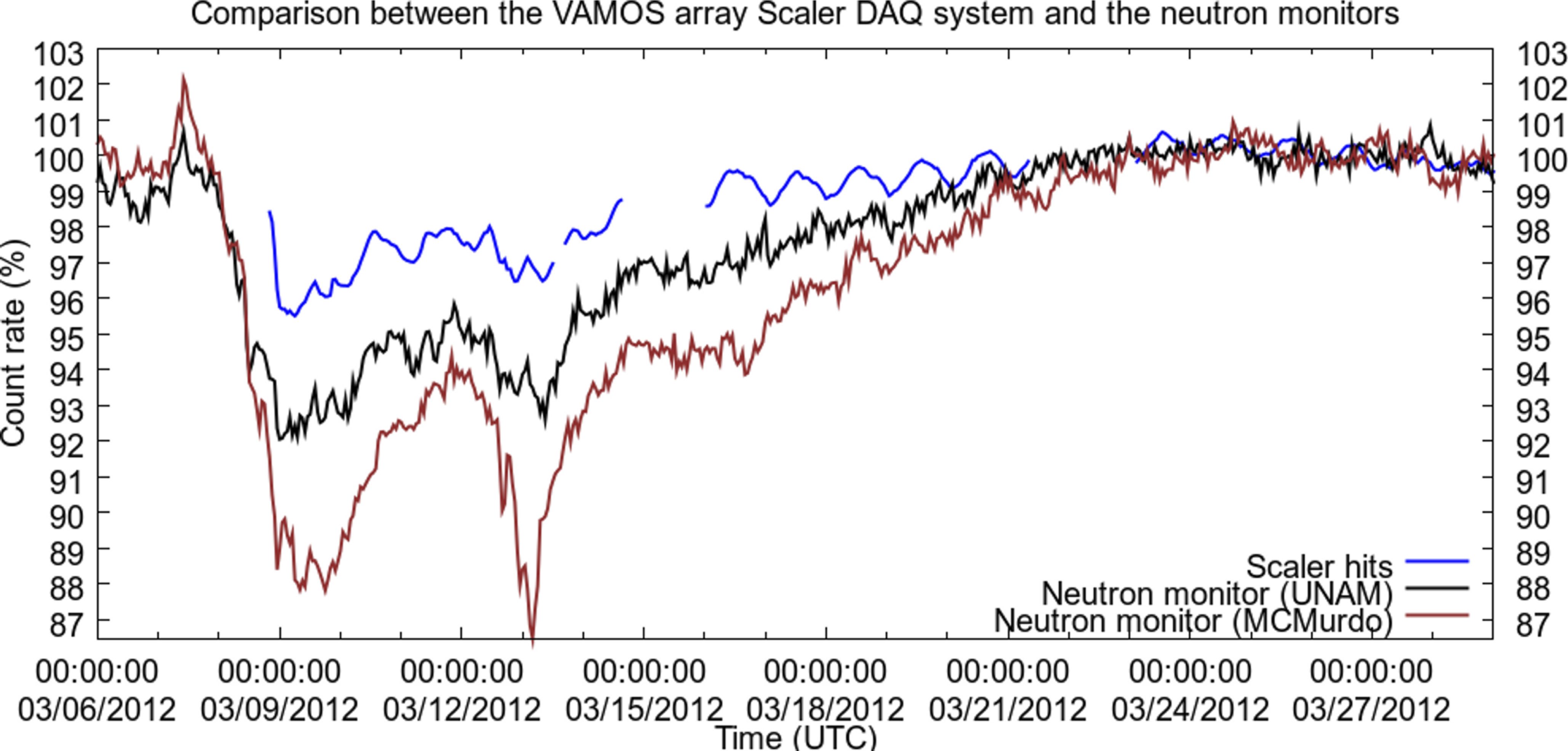}
  \caption{Comparison between the averaged count rates corrected by atmospheric pressure for the Scaler DAQ system at the VAMOS array in blue (magnetic rigidity cutoff 9 GV), the Mexico City Cosmic Rays Observatory in black (magnetic rigidity cutoff 8.2 GV) and the McMurdo station in brown (magnetic rigidity cutoff 0.3 GV).}
 \label{scaler_a}
\end{figure*}

\begin{figure*}[t]
  \centering
  \includegraphics[width=0.57\textwidth]{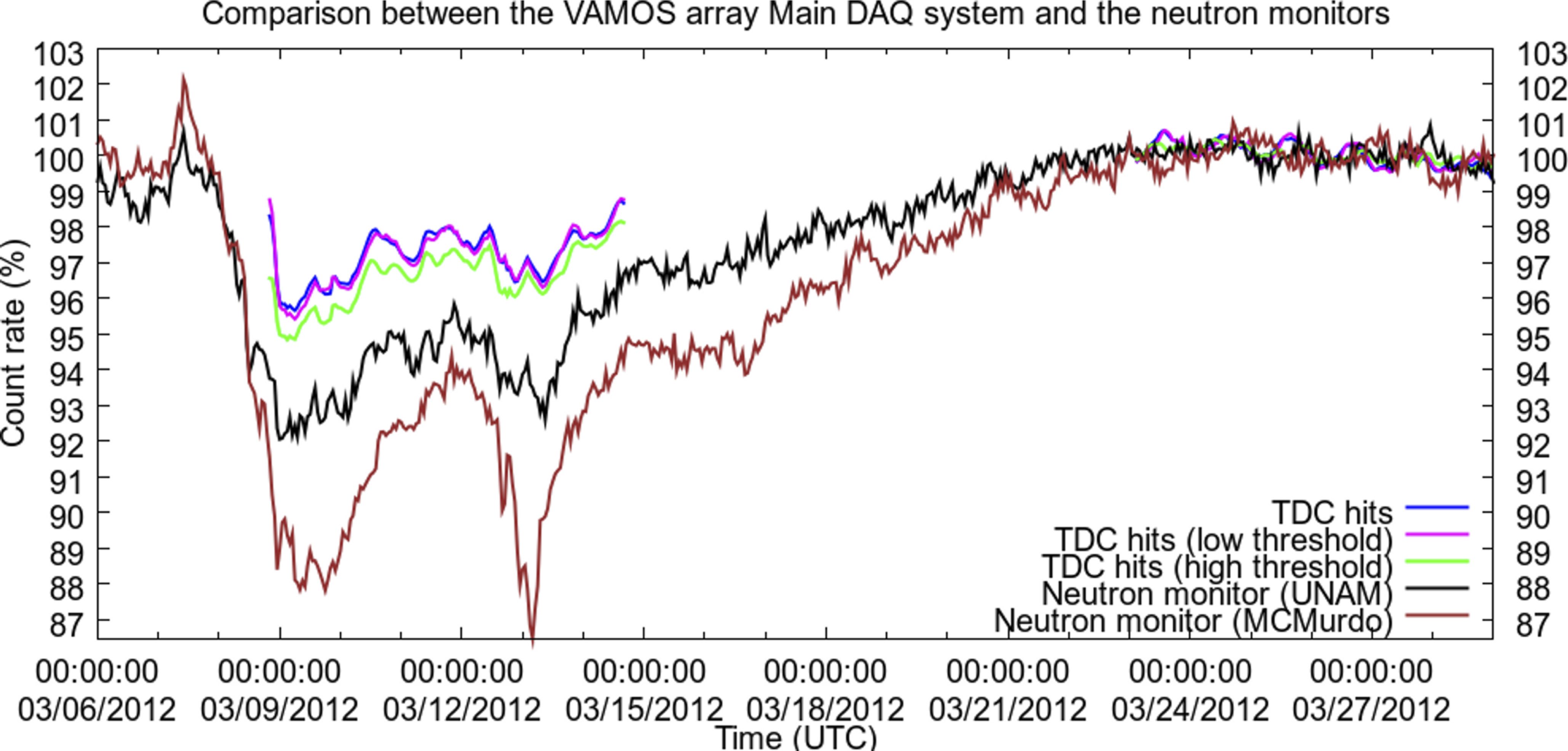}
  \caption{Comparison between the averaged corrected by pressure count rate for the Main DAQ system at the VAMOS array, in blue for the sum of 2-edge and 4-edge hits, in magenta only for 2-edge hits and in green only for 4-edge hits (magnetic rigidity cutoff 9 GV), the Mexico City Cosmic Rays Observatory in black (magnetic rigidity cutoff 8.2 GV) and the McMurdo station in brown (magnetic rigidity cutoff 0.3 GV).}
 \label{tdc_a}
 \end{figure*}

Neutron detectors monitors are routinaly used to monitor Forbush deceases around the world. In particular, the understanding of the modulation of cosmic rays with energies larger than 1 GeV arriving at Earth is significally improved by observations from neutron monitor networks \cite{dasso}.
Two coronal mass ejections (CMEs) propelled reached the Earth on March 7, 2012. The first was traveling faster than 812 km/s while the second with more then 687 km/s.
The results presented in this section include count rates corrected exclusively by atmospheric pressure as done costumarily \cite{dasso}, \cite{icrc2011b}, \cite{icrc2009} and \cite{nima}.
The resulting corrected count rate still shows a residual 12 hour cycle modulation because of they were not corrected by atmospheric temperature given that the temperature sensors were placed incorrectly.
Given that the VAMOS array was not operating before March 8, 2012 we were not able to see the instant when the shock arrived to earth and when the count rate dropped because of the ejecta arrival.
Figures \ref{scaler_a} and \ref{tdc_a} show the results for the averaged signal of 11 PMTs from the two DAQ systems of the VAMOS array, corrected by atmospheric pressure, along with the signals from the neutron monitor located at UNAM in Mexico City and the McMurdo station.
The comparison between the data from the VAMOS array and both neutron monitors is showed in Figures \ref{scaler_a} and \ref{tdc_a} for the Scaler DAQ system and the Main DAQ system, respectively.
Even though there is an excellent agreement in the count rate variations, it is important to keep in mind that the two kinds of experiments detect different kinds of particles because they use different detection principles.
While the VAMOS array detects the hadronic, muonic and the electromagnetic components of the extended atmospheric shower, the neutron monitors (NM64) detect only neutrons from the hadronic component.
One also can see the effect of the magnetic rigidity cutoff on the magnitude of the count rate drop which is different according to the latitude where the experiments are located.
\section*{Conclusions}
We have analyzed data from the VAMOS array of the HAWC Observatory taken during the occurrence of the Forbush decrease that occurred on March 2012. These data  covered a period of 23 days.
There is an excellent similarity between the variations in the count rate detected by the VAMOS array, using the Scaler DAQ system and the Main TDC-based DAQ system, and two neutron detector monitors, one located in Mexico City and the other in the South Pole. 

\section*{Acknowledgement}
We acknowledge the support from: US National Science Foundation (NSF); US Department of Energy Office of High-Energy Physics; The Laboratory Directed Research and Development (LDRD) program of Los Alamos National Laboratory; Consejo Nacional de Ciencia y Tecnolog\'{\i}a (CONACyT), M\'exico; Red de F\'{\i}sica de Altas Energ\'{\i}as, M\'exico; DGAPA-UNAM, M\'exico; and the University of Wisconsin Alumni Research Foundation. 

\clearpage


\end{document}